\documentclass[]{aa}  

\usepackage{graphicx}
\usepackage{txfonts}

\usepackage[utf8]{inputenc}
\usepackage{natbib}
\usepackage{verbatim}
\usepackage{pdfpages}
\usepackage{float}
\usepackage{amsmath}
\usepackage{color}
\usepackage{lipsum}
\usepackage{fancyhdr}
\usepackage{tabularx}
\usepackage{longtable}
\usepackage{url}
\usepackage{hyperref}
\usepackage{breakurl}

\newenvironment{my_itemize}{
	\begin{itemize}
	\setlength{\itemsep}{1pt}
	\setlength{\parskip}{0pt}
	\setlength{\parsep}{0pt}}{\end{itemize}
}

\begin{document}
\bibliographystyle{aa}
\title{Evolution of the mass, size, and star formation rate\\ in high-redshift merging galaxies}
\subtitle{MIRAGE - A new sample of simulations with detailed stellar feedback}

\author{
	V. Perret 
	\inst{1}
	\and F. Renaud
	\inst{2}
	\and B. Epinat
	\inst{1}
	\and P. Amram
	\inst{1}
	\and F. Bournaud
	\inst{2}
	\and T. Contini
	\inst{3,4}
	\and R. Teyssier
	\inst{5}
	\and J.-C. Lambert
	\inst{1}
	}

\institute{
	Aix Marseille Université, CNRS, LAM (Laboratoire d'Astrophysique de Marseille), 13388, Marseille, France 
	\and
	CEA, IRFU, SAp, 91191 Gif-sur-Yvette, France
	\and
	Institut de Recherche en Astrophysique et Planétologie (IRAP), CNRS, 14, avenue Edouard Belin, F-31400 Toulouse, France
	\and 
	Université de Toulouse, UPS-OMP, IRAP, Toulouse, France
	\and
	Institute for Theoretical Physics, University of Zurich, CH-8057 Zurich, Switzerland
	}

\date{}

\abstract
  	{ In $\Lambda$-CDM models, galaxies are thought to grow both through continuous cold gas accretion coming from the cosmic web and episodic merger events. The relative importance of these different mechanisms at different cosmic epochs is nevertheless not yet well understood.}
	{We aim at addressing the questions related to galaxy mass assembly through major and minor wet merging processes in the redshift range $1<z<2$, an epoch corresponding to the peak of the cosmic star formation history. A {significant} fraction of Milky Way-like galaxies are thought to have undergone an unstable clumpy phase at this early stage. We focus on the behavior of the young clumpy disks when galaxies are undergoing gas-rich galaxy {mergers}.}
	{Using the adaptive mesh refinement code \uppercase{Ramses}, we build the Merging and Isolated high-Redshift Adaptive mesh refinement Galaxies (MIRAGE) sample. It is composed of 20 mergers and 3 isolated idealized disks simulations, which sample disk orientations and merger masses. Our simulations can reach a physical resolution of 7 parsecs, and include: star formation, metal line cooling, metallicity advection, and a recent physically-motivated implementation of stellar feedback which encompasses OB-type stars radiative pressure, photo-ionization heating, and supernovae.}	
	{The star formation history of isolated disks shows stochastic star formation rate, which proceeds from the complex behavior of the giant clumps. {Our minor and major gas-rich merger simulations do not trigger starbursts, suggesting a saturation of the star formation due to the detailed accounting of stellar feedback processes in a turbulent and clumpy interstellar medium fed by substantial accretion from the circum-galactic medium.} Our simulations are globally close to the normal regime of the disk-like star formation on a Schmidt-Kennicutt diagram. The mass-size relation and its rate of evolution in the redshift range $1<z<2$ matches observations, suggesting that the inside-out growth mechanisms of the stellar disk do not necessarily {require to be achieved through a cold accretion}.
	}
	{}

\keywords{galaxies: evolution --
                galaxies: formation --
                galaxies: high-redshift --
                galaxies: star formation --
                galaxies: interactions --
                methods: numerical
               }

\maketitle

\section{Introduction}

$\Lambda$-CDM cosmological simulations tend to show that major merger is at play to shape galaxies properties at high-redshift \citep{2009ApJ...702.1005S}. Although it is often set as a competitor of the smooth cold gas accretion along cosmic filaments, which is believed to be very efficient to feed star formation \citep{2009Natur.457..451D, 2009MNRAS.395..160K}, mergers still contribute to around a third of the baryonic mass assembly history \citep{2009ApJ...694..396B,2009ApJ...703..785D}. The pioneering work of \cite{1972ApJ...178..623T} first highlighted that disk galaxy mergers are able to drive large amounts of baryons in tidal tails. \cite{1994ApJ...437L..47M} showed that the redistribution of gas can fuel star formation enhancement in the core of the remnants. Furthermore, stars during a merger event are gravitationally heated and can form spheroids \citep{1996ApJ...471..115B,1996ApJ...464..641M}, underlining a convincing link between the late-type and early-type galaxies of the Hubble sequence. {Boxy slowly rotating ellipticals however probably formed at much higher redshifts through multiple minor mergers or in-situ star formation \citep{2010ApJ...725.2312O,2010ApJ...709..218F,2012ApJ...754..115J}}. The study of stars and gas kinematics is a fair way to detect signatures of merger in the recent history of galaxies \citep{2002MNRAS.333..481B, 2003ApJ...591..791A, 2011MNRAS.416.1654B}, and allows constraining the role of mergers in galaxy mass assembly.
The last decade has seen the first resolved observations of galaxies in the redshift range $0.5<z<3$ using integral fields unit spectrographs (IFU) \citep{2008A&amp;A...477..789Y, 2009ApJ...706.1364F, 2009ApJ...697.2057L, 2011A&amp;A...528A..88G, 2012A&amp;A...539A..91C}, where a peak is observed in the cosmic star formation history. This peak located around $z\sim2$ \citep{2006ApJ...651..142H, 2008A&amp;A...477..789Y} could arise from an intense merger activity, since it is an efficient mechanism to produce starbursts in the local Universe.
\cite{2008ApJ...682..231S} performed a kinematical analysis on the high-z IFU SINS sample to determine the fraction of mergers. To calibrate this analysis, a set of local observations \citep{2006MNRAS.366..812C,2006MNRAS.367..469D,2005MNRAS.360.1201H}, hydrodynamical cosmological simulations \citep{2007ApJ...658..710N} and toy models \citep{2006ApJ...645.1062F} were used. The haloes in the \cite{2007ApJ...658..710N} cosmological simulations were selected to host a merger around $z=2$. Although the baryon accretion history makes these simulations credible in terms of global mass assembly, the resulting low number of haloes could be considered as not sufficient to statistically detect various merger signatures. 

The GalMer database \citep{2010A&amp;A...518A..61C} favors a statistical approach with hundreds of idealized merger simulations, which are probing the orbital configurations. GalMer is relevant for {studying such merger signatures} at low-redshift \citep{2008A&amp;A...492...31D}, however the low gas fractions makes the comparison with high-redshift galaxies impossible. Additionally, simulating the inter-stellar medium (ISM) of high-redshift galaxies requires to correctly resolve the high-redshift disk scaleheights, which can otherwise prevent artificially the expected Jeans instabilities. Indeed, it is now commonly accepted that high-redshift disks are naturally subject to such instabilities \citep{2009ApJ...692...12E,2007ApJ...658..763E}. The high gas fractions at $z>1$ \citep{2010ApJ...713..686D, 2010Natur.463..781T} are strongly suspected to be able to drive violent instabilities which fragment the disks into large star-forming clumps \citep{2008A&amp;A...486..741B} and generate turbulent velocity dispersions \citep[e.g.][]{2012A&amp;A...539A..92E, 2008ApJ...680..246T}. Therefore, the canonical image of smooth extended tidal tails falling onto the merger remnant cannot be valid in the context of gas-rich interactions \citep{2011ApJ...730....4B}. 

The ability to form such clumps is important to understand the complex behavior of high-redshift galaxies, but it is also essential to prevent the over-consumption of gas expected at these very high gas densities from the  classical Schmidt law. In order to match the Kennicutt-Schmidt (KS) relation \citep{1998ApJ...498..541K} and to have acceptable gas consumption timescales, an efficient stellar feedback is required to deplete the gas reservoir of the star-forming regions. Indeed, cosmological simulations with no or weak feedback models produce galaxies with too many baryons in the galactic plane \citep{2009MNRAS.396.2332K} when compared to the abundances matching techniques \citep{2010MNRAS.404.1111G}. The constraints on the inter-galactic medium (IGM) metal enrichment also imply that baryons entered at some points into galaxies and underwent star formation \citep{2001ApJ...561..521A}. It has been demonstrated that scaling supernovae stellar winds in cosmological simulations to the inverse of the mass of the host galaxy produces models in reasonable agreement with the local mass function \citep{2010AAS...21537603O}. {It is therefore essential to constrain the parameters controlling the stellar feedback processes in order to better understand the scenarios of galaxy evolution.}

To get an insight into the various processes of galaxy mass assembly, such as mergers, the Mass Assembly Survey with SINFONI in VVDS (MASSIV, \citealt{2012A&amp;A...539A..91C}) aims at probing the kinematical and chemical properties of a significant and representative sample of high-redshift ($0.9 < z < 1.8$) star-forming galaxies. 
Observed with the SINFONI integral-field spectrograph at the VLT and built upon a simple selection function, the MASSIV sample provides a set of 84 representatives of normal star-forming galaxies with SFRs ranging from 5 to 400 $M_{\odot}.yr^{-1}$ in the stellar mass regime $10^9-10^{11} M_{\odot}$.
Compared to other existing high-z IFU surveys, the main advantages of the MASSIV sample are its representativeness as it is flux-selected from the magnitude-limited VVDS survey \citep{2005A&amp;A...439..845L}, and its size that allows to probe different mass and SFR ranges, while keeping enough statistics in each category. Together with the size of the sample, the spatially-resolved data therefore allows to discuss global, volume averaged, galaxy kinematic, and chemical properties across the full mass and SFR ranges of the survey to derive robust conclusions for galaxy mass assembly on cosmological timescales.
By studying strong kinematic signatures of merging and detecting pairs in the first epoch MASSIV, \cite{2012A&amp;A...539A..92E} have shown that the fraction of interacting galaxies is up to at least one third of the sample and that more of a third of the galaxies are non-rotating objects. 
In addition, the fraction of non-rotating objects in mergers is higher than in isolated galaxies. This suggests that a significant fraction of isolated non-rotating objects could be mergers in a transient state in which the gas is not dynamically stable. 
Furthermore, based on the whole MASSIV sample, \cite{2013A&amp;A...553A..78L} found a gas-rich major merger fraction of $\sim$20\% in the redshift range $1<z<1.8$ and a gas-rich major merger rate of $\sim$0.12.
The quantification of kinematical signatures of interacting galaxies and mergers and the understanding of the high fraction of non-rotating systems, the existence of inverse metallicity gradient in some disks \citep{2012A&amp;A...539A..93Q}, and more generally, a comprehensive view of the process of formation of turbulent and clumpy gaseous galaxy disks have motivated the building of a set of simulations of merging galaxies in the redshift range probed with MASSIV, i.e. the MIRAGE simulations.

We describe in this paper, a set of 20 idealized galaxy mergers and three isolated disks using adaptive mesh refinement (AMR) simulations with a physically-motivated implementation of stellar feedback\footnote{Movies of the simulations of the present paper are available at:\\ \tiny{\url{http://www.youtube.com/playlist?list=PL_oPMhue14ZSyxcuFiJrUXI-6ej8Q7rv7}}}. 
This paper focuses on the presentation of the MIRAGE sample, the numerical technique employed, and the global physical properties deduced. The analysis is extended in a companion paper \cite{2013arXiv1307.7136B} presenting a study of the clumps properties in the three isolated disk simulations of the MIRAGE sample. 
The paper is organized as follows.
In section \ref{simus}, we describe the numerical technique used to build our simulations sample.
In section \ref{dice}, we specifically describe the idealized initial conditions generation. For this purpose we introduce the new public code DICE and summarize the different numerical techniques used to generate stable galaxies models.
Section \ref{definition} reviews the MIRAGE sample definition of galactic models and orbital parameters.
Section \ref{properties} describes the global properties of the sample. We present the star formation histories, the disk scalelengths evolution, and its position on the KS relation.

\section{Simulations}
\label{simus}

We have run a set of idealized AMR high-redshift galaxy simulations. The sample encompasses 20 major/minor galaxy mergers and three isolated disks, with a high gas fraction {(>50\%)} typical of $1<z<2$ galaxies \citep{2010ApJ...713..686D}, evolved over 800 Myr. In this work, we choose to balance the available computational time between high resolution and statistical sampling of the orbital parameters to provide new insights on the galactic mass assembly paradigm. 

\subsection{Numerical Technique}
\label{num_tech}

In order to build our numerical merger sample, we use the adaptive mesh refinement (AMR) code \verb|RAMSES| \citep{2002A&amp;A...385..337T}. The time integration of the dark matter and the stellar component is performed using a particle-mesh (PM) solver, while the gas component evolution is insured by a second-order Godunov integration scheme. The code has proven its ability to model the complexity of interstellar gas on various galaxies simulations (e.g. \citealt{2008A&amp;A...477...79D, 2013MNRAS.429.3068T}).
The computational domain of our simulations is a cube with a side $l_{box}$=240 kpc, and the coarsest level of the AMR grid is $\ell_{min}=7$, which corresponds to a cartesian grid with $(2^7)^3$ elements and with a cell size of {$\Delta x = 1.88$ kpc}. The finest AMR cells reach the level $\ell_{max}=15$, where the cell size corresponds to $\Delta x=7.3$ pc. The grid resolution is adapted at each coarse time step between the low refinement level ($\ell_{min}=7$) and the high refinement level ($\ell_{max}=15$). Each AMR cell is divided into eight new cells if at least one of the following assertions is true: (i) it contains a gas mass greater than $1.5\times 10^4 M_{\odot}$ (ii) it contains more than 25 particles (dark matter or stars) (iii) the local Jeans length is smaller than 4 times the current cell size. This quasi-Lagrangian refinement scheme is comparable to the one introduced in \cite{2010ApJ...720L.149T} and  \cite{2010MNRAS.409.1088B}.

The star formation is modeled with a Schmidt law triggered when the density $\rho_{gas}$ overcomes the threshold $\rho_0$=100 cm$^{-3}$, with an efficiency $\epsilon_{\star}$=1\%:

\begin{equation}
\dot{\rho}_{\star}=
	\left\lbrace
	\begin{array}{ccc}
		0  & \mbox{if} & \rho_{gas}<\rho_0 \\
		0  & \mbox{if} & T>2\times10^5 \mathrm{K}\\
		\epsilon_{\star}\rho_{gas}/t_{ff}   & \mbox{else},
	\end{array}\right.
\end{equation}
where $\dot{\rho}_{\star}$ is the local star formation rate, $t_{ff}=\sqrt{3\pi /(32G\rho_{gas})}$ is the free-fall time computed at the gas density $\rho_{gas}$, and $T$ is the temperature of the cell considered.  AMR cells with temperature greater than $2\times10^5$K are not allowed to form stars.

The gravitational potential is computed using a PM scheme with a maximum level $\ell_{max,part}=13$ for the grid, which ensures a gravitational softening of at least ~29 pc for Lagrangian particles. This choice prevents a low number of dark matter particles per cell, often synonym of N-body relaxation.
We use a thermodynamical model modeling gas cooling provided by the detailed balance between atomic fine structure cooling and UV radiation heating from a standard cosmic radiation background by using tabulated cooling and heating rates from \cite{2004A&amp;A...416..875C}. In this model, the gas metallicity acts like a scale factor on the cooling rate. 

\begin{figure}[htbp]
	\centering
	\includegraphics[width=9cm]{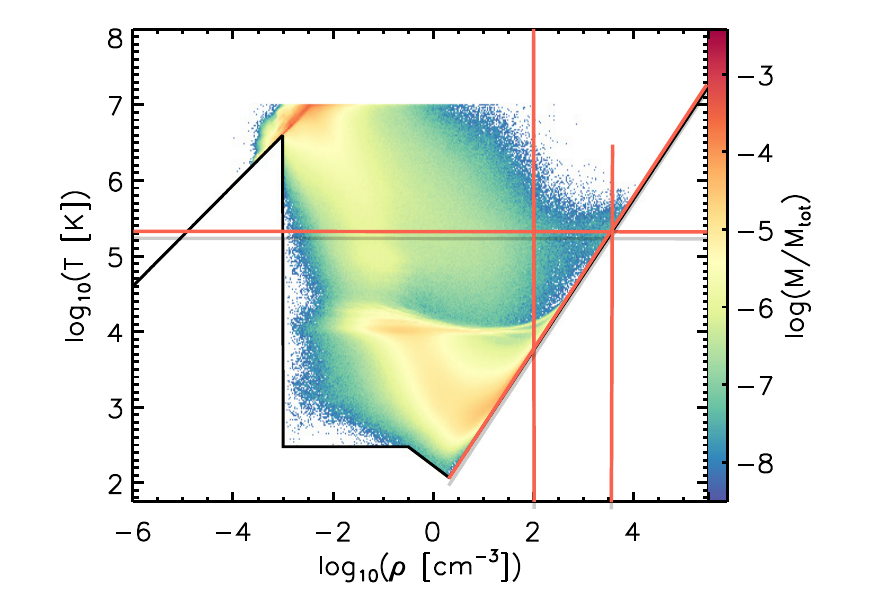}
	\caption{Density-temperature diagram of the G1 model integrated over 800 Myr {(see section \ref{galaxy_models} for a description of the galaxy models)}. The black line represents the temperature floor described in section \ref{num_tech}. Each AMR cell contribution to the 2D histogram is weighted by its mass. $M$ represents the gas mass contribution to a bin in the $\rho-T$ plane, color coded on a log-scale. $M_{tot}$ is the total gas mass, used as a normalization factor.}
	\label{G1_DT_diagram}
\end{figure}

The gas is forced to stay within a specific area in the density-temperature diagram in order to prevent multiple numerical artifacts (see Fig. \ref{G1_DT_diagram}):
\begin{itemize}
\item In the low IGM density regime ($\rho<10^{-3}$ cm$^{-3}$), we ensure a gravo-thermal equilibrium for the gas by introducing a temperature floor in the halo following a gamma polytrope at the virial temperature $T_{min}(\rho)=4 \times 10^6 (\rho /10^{-3})^{2/3}$ K, as in \cite{2010MNRAS.409.1088B}. 

\item For densities between $10^{-3}$ cm$^{-3}$$<\rho<0.3$ cm$^{-3}$ the temperature floor is isothermal and set to $T_{min}(\rho)=T_{floor}$. At full resolution (i.e. $\ell_{max}=15$), we have $T_{floor} = 300$ K. {The densest IGM can reach the $\rho=10^{-3} \mathrm{cm}^{-3}$ limit, and can condense on the gaseous disk.}

\item For densities above 0.3 cm$^{-3}$ we use the temperature floor $T_{min}(\rho)=300\times (\rho/0.3)^{-1/2}$ K; this choice allows us to have a dynamical range in the thermal treatment of the gas up to 30 times colder than {the slope of the thermodynamical model used in \cite{2010ApJ...720L.149T} and \cite{2010MNRAS.409.1088B}}.

\item A density-dependent pressure floor is implemented to ensure that the local Jeans length is resolved at least by $n_{Jeans}=6$ cells in order to avoid numerical fragmentation, as initially proposed by \cite{1997ApJ...489L.179T}.  
This Jeans polytrope acts like a temperature floor for the very dense gas: it overcomes the cooling regime of the temperature floor starting from $\rho=2.6$ cm$^{-3}$ when the resolution is maximum, i.e. cells with a size of 7.3 pc. {The Jeans polytrope is described by the equation: $T_{min}(\rho) = \rho G m_H (l_{box} n_{Jeans} / 2^{\ell_{max}})^2 / (\pi k_B \sqrt{32})$ with $m_H$ the proton mass and $k_B$ the Boltzmann constant.}

\item We impose a maximum temperature for the gas $T_{max}=10^7$ K. Indeed, the clumps generated by Jeans instabilities typical of gas-rich disks \citep{2008A&amp;A...486..741B} lead to regions of low density inside the disk, where supernovae can explode. This thermal explosion is thus able to produce sound speed greater than 1000 km.s$^{-1}$, which affects the time step in the Godunov solver. Setting an upper limit to the temperature is not fully conservative in terms of energy, but our choice of $T_{max}$ ensures a viable time step and a reasonably low energy loss. This issue typical of grid codes is handled in the same way in the recent work of \cite{2013arXiv1309.2942H}.
\end{itemize}

{Due to non-periodic boundary conditions applied to the AMR box, we impose a zero density gradient in the hydrodynamical solver at the boundaries. To avoid galaxies passing close to the edges of the box which could induce numerical artifacts, it is required to have sufficiently large AMR volume to encompass the whole trajectories of both galaxies through the simulation duration. }

\subsection{Feedback models}
Because we do not resolve individual stars, each stellar particle models a population that contains massive OB-type stars with masses $M>4M_{\odot}$ \citep{2012arXiv1202.0791P} and which are responsible for injecting energy into the surrounding ISM. Assuming a \cite{1955ApJ...121..161S} initial mass function, we consider that a fraction $\eta = 20\%$ of the mass of a stellar particle contributes to stellar feedback, which is effective during 10 Myr after the star particle is spawned. 
We use the \cite{2013MNRAS.tmp.2414R} {physically-motivated} model implementation for the OB-type stars feedback, and summarize the three main recipes below: 
\begin{my_itemize}
	\item {Photo-ionization}: OB-type stars produce highly energetic photons capable of ionizing the surrounding ISM. Using a simple model for the luminosity of the star, the radius of the \cite{1939ApJ....89..526S} sphere is computed according to the mean density of electrons $n_e$. The gas temperature inside the HII regions is replaced by an isothermal branch at $T_{HII}=10^4$ K. The radius of the HII sphere is computed via the equation:
	\begin{equation}
		R_{HII} = \frac{3}{4\pi} \frac{L_*}{n_e^2 \alpha_r} \;,
	\end{equation}
	where $L_*$ is the time dependent luminosity of the star in terms of ionizing photons, and $\alpha_r$ is the effective recombination rate.

	\item {Radiative pressure}: Inside each HII bubble, a kinetic momentum $\Delta\mathrm{v}$ is distributed as a radial velocity kick over the time interval $\Delta t$, matching the time step of the coarsest level of the simulation. This velocity kick is computed using ionizing photons momentum, which is considered to be transferred to the gas being ionized, i.e. the gas within the radius $R_{HII}$ of the Strömgren sphere. 
	\vspace{0.5cm}
	\begin{equation}
		\Delta \mathrm{v} = k \frac{L_*h\nu}{M_{HII}\mathrm{v_c}}\Delta t,
	\end{equation}
	with $h$ the Planck constant, $\mathrm{v_c}$ the speed of light, $M_{HII}$ the gas mass of the bubble affected by the kick, and $\nu$ the frequency of the flux representative of the most energetic part of the spectrum of the ionizing source. In this model, it is considered that Lyman-$\alpha$ photons dominate this spectrum, implying $\nu = 2.45 \times 10^{15}$ $s^{-1}$. The distribution of the momentum carried by ionizing photons is modeled by the trapping parameter $k=5$, which basically counts the number of diffusion per ionizing photon and energy loss. This value may appear to be rather high compared to recent works (e.g. \citealt{2012ApJ...760..155K}), but is more acceptable once considered that we miss other sources of momentum such as proto-stellar jets and stellar winds \citep{2013MNRAS.432..455D}.

	\item {Supernova explosions}: 
		We follow the implementation of supernova feedback of \cite{2008A&amp;A...477...79D}: the OB-type star population which reaches 10 Myr transforms into supernovae (SNe) and releases energy, mass and metals into the nearest gas cell. The gas which surrounds the supernovae receives a fraction $\eta=20\%$ of the stellar particle mass as well as a specific energy $E_{\mathrm{SN}}=2\times10^{51}$ ergs/$10M_{\odot}$, which is the product of the thermo-nuclear reactions. {The energy injected by each SN is higher by a factor of two compared to some other works \citep[e.g.][]{2013MNRAS.429.3068T, 2012MNRAS.420.2662D}, but simulations of individual type II SN releasing such energy could be frequent in the early Universe \citep{2010ApJ...709...11J}. Moreover, the use of an IMF with a lower statistical contribution of low mass stars would imply higher values for $\eta$ (e.g. $\eta\simeq35\%$ for \citealt{2001MNRAS.322..231K} IMF), which balances our choice of a high value for $E_{\mathrm{SN}}$.}
Each supernovae event also releases into the surrounding ISM metals derived form the nucleosynthesis following the equation:

\begin{equation}
	Z=y+(1-y)  Z_{ini},
\end{equation}	

with $Z$ the mass fraction of metals in the gas, $Z_{ini}$ the initial metal fraction of the supernova host, and $y$ the yield which is set to $y=0.1$, as in \cite{2012MNRAS.420.2662D}.	

In order to account for non-thermal processes due to gas turbulence at sub-parsec scales we follow the revised feedback prescriptions of \cite{2013MNRAS.429.3068T}.
	The numerical implementation is similar to introducing a delayed cooling in the Sedov blast wave solution. At each coarse time step, the fraction of gas released by SNe in AMR cells is evaluated in a passively advected scalar; the gas metal line cooling is switched off as long as:
	\begin{equation}
		\frac{m_{ejecta}}{m} > 10^{-3},
	\end{equation} 

	with $m$ the gas mass of the cell, and $m_{ejecta}$ the total mass of the gas ejected by SNe in the same cell.
	In order to model the turbulence dissipation, the mass of gas contributing to the Sedov blast wave is lowered by a factor $\gamma$ at each coarse time step following:
	\begin{equation}
		\gamma = \exp\left(-\frac{dt_{cool}}{t_{dissip}}\right),
		\label{gamma}
	\end{equation} 
	 with $dt_{cool}$ the cooling time step, and $t_{dissip}$ a typical timescale for the turbulence induced by the detonation.
	 The dissipation timescale for the unresolved sub-grid turbulent structures is the crossing time \citep{1999ApJ...524..169M}, i.e the ratio of the numerical resolution over the velocity dispersion.
	 Since our simulations are able to resolve structures down to 7 pc, we presume that the non-thermal velocity dispersion in the smallest AMR cells is close to 5 km.s$^{-1}$, which is a typical supersonic speed in regions of star formation with gas temperatures below $10^3$ K \citep{2012A&amp;ARv..20...55H}. Under these assumptions, we set $t_{dissip}=$2 Myr.

\end{my_itemize}

{
Our feedback model does not assume the systematic destruction of the clumps by the star formation bursts following  their formation, contrary to what is done in some other works \citep[e.g.][]{2013MNRAS.tmp..733H, 2012ApJ...745...11G}.}
Smaller clumps are subject to disruption, but larger clumps may survive such thermal energy injection. This model clearly favors the scenario of long-lived star-forming clumps, which we aim to address in this study. 

\section{DICE: a new environment for building disk initial conditions}
\label{dice}
The initial conditions of the MIRAGE sample are constructed using a software developed for the purpose of the task, named Disk Initial Conditions Environment (DICE). DICE is an implementation of the numerical methods described in \cite{2005ApJ...620L..79S}. It is able to setup multiple idealized galaxies in a user friendly context. The software is open source and available online\footnote{\url{http://code.google.com/p/dice-project/}}.

\subsection{Density distributions}
DICE initial conditions are generated using Lagragian particles whose distributions are built using a Metropolis-Hasting Monte-Carlo Markov Chain algorithm \citep{1953JChPh..21.1087M}. The strength of this algorithm lies in its ability to build a distribution for a sample of Lagrangian particles having only the knowledge of the probability distribution function. 
{After having initialized the first Lagrangian particles of each component (disk, bulge, gas, halo, etc.) to a probable location, the algorithm loops over the desired number of Lagrangian particles and iteratively produces a candidate position for each of them. The probability to set a Lagrangian particle to the randomly picked candidate cartesian position $\mathbf{x}'$ depends on the cartesian position $\mathbf{x}$ of the previous particle in the loop, and is written:}

\begin{equation}
	\alpha(\mathbf{x},\mathbf{x}')=\min \left(1, \frac{\rho(\mathbf{x})P(\mathbf{x},\mathbf{x}')}{\rho(\mathbf{x}')P(\mathbf{x}',\mathbf{x})}  \right),
\end{equation}

{with $\rho(\mathbf{x})$ the density function of the considered component at the position $\mathbf{x}$, and $P(\mathbf{x},\mathbf{x}')$ the probability to place the particle at $\mathbf{x}'$ considering the position $\mathbf{x}$ of the previous particle.
Indeed, our implementation uses a Gaussian walk, meaning that the candidate coordinates are generated using the rule:}

\begin{equation}
	\mathbf{x}'=\mathbf{x}+\sigma W,
\end{equation}

{where $W$ is a standard Gaussian random variable, and $\sigma$ a dispersion factor tuned to a fixed fraction of the targeted scalelength of the component to build, ensuring a satisfying convergence. For each particle, a uniform random value $\tau \in [0,1]$ is picked, and the position of the Lagrangian particle is set to $\mathbf{x}'$ if $\tau\leq\alpha$, and is set to $\mathbf{x}$ otherwise.}
The first 5\% of the iterations to build the distribution are not taken into account because they are considered as a ``burning period'' to account for eventual poor choice of initial values.

To fit the system in the finite AMR domain, we cut the density profiles of all the components. We apply these cuts using an exponential {truncation} profile {at the edges of each component}, in order to prevent strong discontinuities nearly the cut region which would make the numerical differentiation quite unstable. {The scalelength of the exponential {truncation} profile is set to be one percent of the gas disk scaleheight.}

\subsection{Gravitational potential}
	In order to setup the velocities in our initial conditions, we compute the gravitational potential using a PM technique. We first interpolate the densities of all the components onto a cartesian grid using a cloud-in-cell scheme. We compute the gravitational potential $\Phi$ by solving the Poisson equation:
	
	\begin{equation}
		\Phi(\mathbf{x})=\int \mathcal{G}(\mathbf{x},\mathbf{x}')4\pi \rho(\mathbf{x}')d^3\mathbf{x}'\;,
	\end{equation}
	where $\mathcal{G}$ is the Green function, $\rho$ is the density function of all the mass components interpolated on the cartesian grid.
	We compute this integral by performing a simple product in the Fourier plane, which is equivalent to a convolution in the real plane. We eliminate the periodicity associated to the fast Fourier transform algorithm using the zero-padding technique described in \cite{1988csup.book.....H}.
	
	\subsection{Velocities}
	In order to fully describe our system, we assume that the mean radial and vertical velocities $\langle v_r \rangle$ and $\langle v_z \rangle$ are equal to zero. 
	The velocities of each Lagrangian particle are determined by integrating the Jeans equations \citep{2009PhT....62e..56B}, assuming that the velocity distribution is shaped as a tri-axial Gaussian.
	For the dark matter halo and the stellar bulge, we numerically solve the equations:
	\begin{equation}
		\langle v_z^2 \rangle = \langle v_r^2 \rangle = \frac{1}{\rho}\int_z^\infty \rho(r,z') \frac{\partial \Phi}{\partial z'}dz' \;,
	\end{equation}
	
	\begin{equation}
		\langle v_{\phi}^2 \rangle = \langle v_r^2 \rangle + \frac{r}{\rho} \frac{\partial(\rho)\langle v_r^2 \rangle}{\partial r} + r\frac{\partial \Phi}{\partial r}\;.
	\end{equation}
	The velocity dispersion can thus be computed using the relation:
	\begin{equation}
		\sigma_{\phi}^2 = \langle v_{\phi}^2 \rangle - \langle v_{\phi} \rangle^2 \; .
	\end{equation}
	The dark matter halo is generally described with an angular momentum which is not specified by the Jeans equations. The streaming component is set to be a small fraction $f_s$ of the circular velocity, i.e. $\langle v_{\phi} \rangle=f_s v_c$. The fraction $f_s$ depends on the halo spin parameter $\lambda$ and the halo concentration parameter $c$ \citep{1999MNRAS.307..162S}, which are used as input parameters in our implementation.
	
	For the stellar disk, we choose to use the axisymmetric drift approximation \citep{2009PhT....62e..56B} which allows fast computation, {although} we caution the risk of using this approximation with thick and dispersion supported disks\footnote{The axisymmetric drift approximation is valid for relatively thin disk. The use of this approximation for thicker disks supported by velocity dispersions might generates relaxation effects which would make the initial conditions unstable.}. This approximation relates the radial Gaussian dispersion to the azimuthal one:
	
	\begin{equation}
		\sigma_{\phi}^2 = \frac{\sigma_r^2}{\eta^2} \;,
	\end{equation}
	with
	\begin{equation}
		\eta^2 = \frac{4}{r} \frac{\partial \Phi}{\partial r} \left( \frac{3}{r}\frac{\partial \Phi}{\partial r} + \frac{\partial^2 \Phi}{\partial r^2} \right)^{-1}.
	\end{equation}

	The Toomre parameter for the stellar disk is written: 

	\begin{equation}
		Q_{stars}=\frac{\sigma_z \kappa}{3.36 G \Sigma},
	\end{equation}
	where $\kappa$  is the so-called epicyclic frequency, and $\Sigma$ is the combined surface density of the stellar and gaseous disks.
	It is used to control the stability of the stellar disk by setting a minimum value for the velocity dispersion $\sigma_z$ which prevents the local Toomre parameter to go below a given limit of 1.5 {in the initial conditions of our simulations, although this parametrization can not prevent the natural fragmentation of the gaseous disk at later stages.}	

	The only component to specify for the gas is the azimuthal streaming velocity, derived from the Euler equation:

	\begin{equation}
		\langle v_{\phi,gas} \rangle  = r \left( \frac{\partial \Phi}{\partial r}+ \frac{1}{\rho_{gas}} \frac{\partial P}{\partial r} \right),
	\end{equation}
	where $P$ is the gas pressure. 
	
{\subsection{Keplerian trajectories}}

	DICE is also able to setup the Keplerian trajectories of two galaxies involved in an encounter. Using the reduced particle approach we can setup the position of the two galaxies with only three input parameters: (i) the initial distance between the two galaxies $r_{ini}$, (ii) the pericentral distance $r_{peri}$ i.e. the distance between the two galaxies when they reach the periapsis of the Keplerian orbit, (iii) the eccentricity of the trajectories, which are equal for both of the galaxies. The position of the barycenter of each galaxy in the orbital plane can be expressed in cartesian coordinates as follow:
	\begin{equation}
	\begin{array}{ll}
		x_{1}=r_{1}\cos(\psi_{1})\;, &  y_{1}=r_{1}\sin(\psi_{1})\;,\\
		\\
		x_{2}=-r_{2}\cos(\psi_{2})\;, & y_{2}=-r_{2}\sin(\psi_{2})\;,
	\end{array}
	\end{equation}
	with $\psi_{1}$ and $\psi_{2}$ respectively the true anomaly of the first and second galaxy. 
	The cartesian velocities $v_x$, $v_y$ of the two galaxies in the orbital plane are computed using: 

	\begin{equation}
	\begin{array}{ll}
		v_{x,1} =  k_1\sqrt{\frac{\gamma}{\mathcal{L}}} \sin(\psi_1)\;, &  
		v_{y,1} = -k_1\sqrt{\frac{\gamma}{\mathcal{L}}} \left[ e+\cos(\psi_1)\right]\;, \\
		\\
		v_{x,2} = -k_2\sqrt{\frac{\gamma}{\mathcal{L}}} \sin(\psi_2)\;, &
		v_{y,2} =  k_2\sqrt{\frac{\gamma}{\mathcal{L}}} \left[e+\cos(\psi_2)\right]\;,
	\end{array}
	\end{equation}
	with $k_i$ the mass fraction of the $i-$galaxy compared to the total mass of the system, $\gamma$ the standard gravitational parameter, $\mathcal{L}$ the semi-latus rectum of the reduced particle of the system, $e$ the eccentricity of the orbits. With these definitions, it is possible to set trajectories of any eccentricity. {This parametrization holds for point mass particles, while galaxies are extended objects which undergo dynamical friction. The galaxies quickly deviates from their initial trajectories because of the transfer of orbital energy towards the energy of each galaxy, which can lead to coalescence.}

\vspace{0.5cm}
\section{Sample definition}
\label{definition}
\subsection{Galaxy models}
\label{galaxy_models}
The different parameters of our disk initial conditions are summarized in Table \ref{properties_simulations}.
We set up three idealized galaxy models based on the MASSIV sample stellar mass histogram \citep{2012A&amp;A...539A..91C}. The choice of the initial stellar masses of our simulations is carried out in order to sample this histogram with all the available snapshots, i.e. in the redshift range $1<z<2$.
{We choose to build our sample out of three disk models with the respective stellar masses: $\log(M_{\star}/M_{\odot})=9.8$ for our low mass disk, $\log(M_{\star}/M_{\odot})=10.2$ for our intermediate mass disk, and $\log(M_{\star}/M_{\odot})=10.6$ for our high mass disk.}
All of our models have a stellar disk and a gaseous disk with an initial gas fraction $f_g$=65\%. The stellar density profile is written: 
	\begin{equation}
		\rho_{stars}(r,z) = \frac{M_{stars}}{2\pi h_{stars}^2} \exp\left(-\frac{r}{r_{stars}}\right) \exp\left(-\frac{z}{h_{stars}}\right),
	\end{equation}

	with $r_{stars}$  the scalelength of the stellar disk, $h_{stars}$ the scaleheight of the stellar disk, and $M_{stars}$ is the un-cutted stellar disk mass. We use the exact same exponential profile to set up the gaseous disk, {with scalelengths 1.68 times shorter than the stellar counterpart as measured in the MASSIV sample data \citep{2012A&amp;A...546A.118V}.}
		We initialize the metallicity in the gas cells modeling the ISM of the disks following an exponential profile to be consistent with the previous prescriptions:

	\begin{equation}
		Z(r) = Z_{core}\exp\left(-\frac{r}{r_{metal}}\right).
	\end{equation}
	We choose to have negative initial metallicity gradients, with values of $r_{metal}$ equal to the gaseous disk scalelength. The fraction of metals in the center $Z(r=0)=Z_{core}$ of each model is chosen to follow the mass-metallicity relation at z=2 found in \cite{2006ApJ...644..813E}. Such a choice combined with the exponential profile provides global metallicities 50 percent lower than the mass-metallicity relation at $z=2$ for starburst galaxies, but this choice is consistent with our aim of modeling normal star-forming galaxies. The numerical implementation of metallicity treatment of the stellar particles ignores the stars present in the initial conditions. It is therefore not required to set a metallicity profile {for these stars.}
	
	Dark matter haloes are modeled using a \cite{1990ApJ...356..359H}  profile, with a spin parameter set close to the conservative value with $\lambda=0.05$ \citep{1992ApJ...399..405W, 1998MNRAS.295..319M}:
	\begin{eqnarray}
		\rho_{halo}(r) = \frac{M_{halo}}{2\pi}\frac{a}{r(r+a)^3}  \;, \\
		a=r_{halo}\sqrt{2 \left( \ln(1+c) - \frac{c}{1+c} \right)}  \;,
	\end{eqnarray}
	where $M_{halo}$ is the total dark matter mass, $a$ is the halo scalelength and $r_{halo}$ is the scalelength for an equivalent \citet*{1997ApJ...490..493N} halo with the same dark matter mass within $r_{200}$ \citep{2005MNRAS.361..776S}. We can therefore define our halo with the frequently used concentration parameter $c$, which is set to a value $c=5$ as it has been measured at $z\sim2$ in N-body cosmological simulations \citep{2001MNRAS.321..559B}. {We do not take into account the mass dependence of the halo concentration function to ensure that our simulations are comparable in terms of disk instability between each other.}	

	Finally, a bulge enclosing 8\% of the total initial stellar mass is modeled using again an Hernquist profile, with a scalelength set to be equal to 20\% of the stellar disk scalelength.

\begin{figure}[htbp]
	\centering
	\includegraphics[width=9cm]{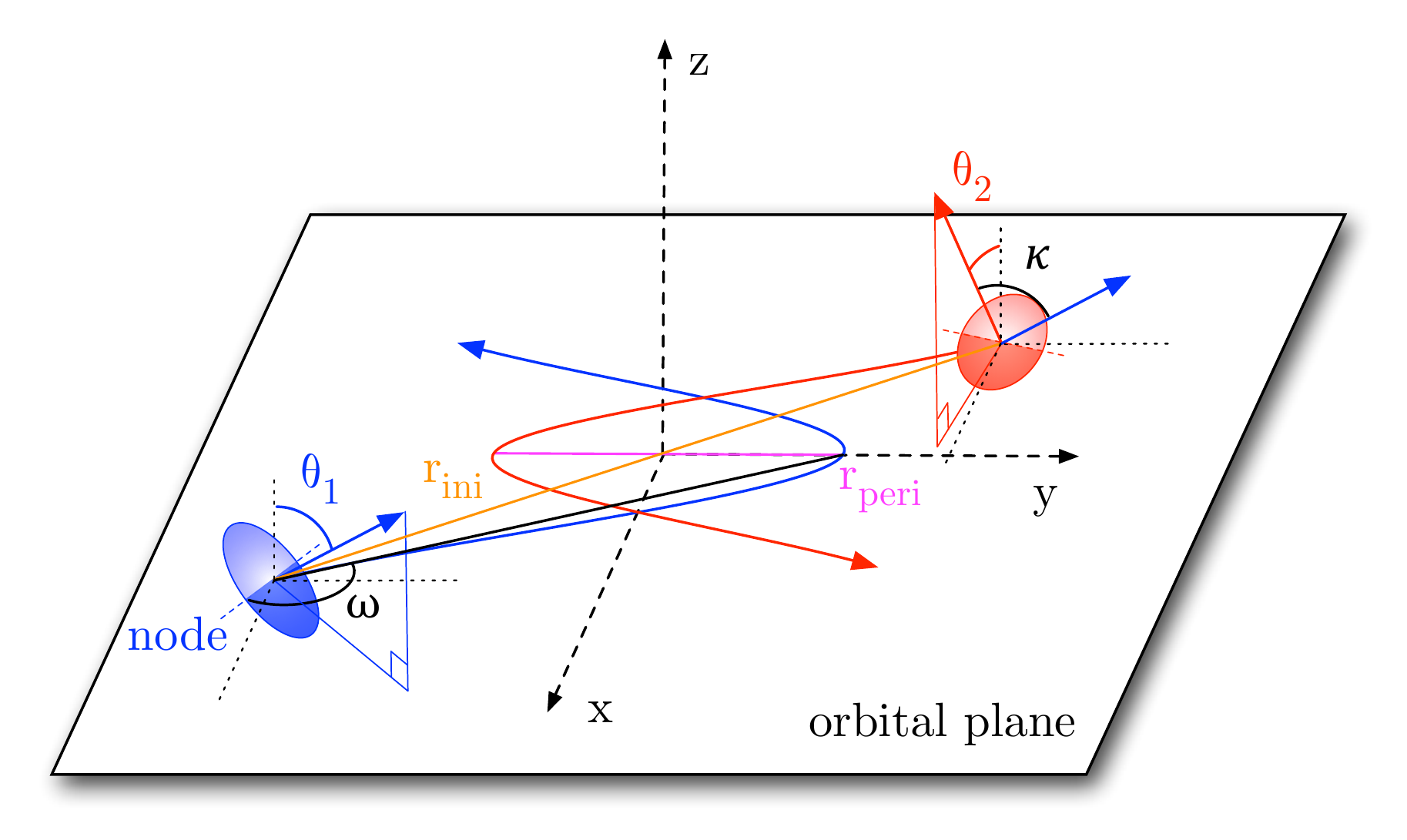}
	\caption{Orbital geometry used in our simulation sample. Four angles define the geometry of the interaction: $\theta_1$, $\theta_2$, $\kappa$, and $\omega$. {The pericentric argument $\omega$, is defined as the angle between the line of nodes (intersection between the orbital plane and the galactic plane) and separation vector at pericenter (black line). The blue/red arrows display the spin orientation for the first/second galaxy.} The blue/red curves represent the trajectory of the first/second galaxy in the orbital plane (x,y). {The centers of the two galaxies are also lying in the orbital plane. The darkest parts of the disks are lying under the orbital plane}.}
	\label{simus_geometry}
\end{figure}

\begin{table}[h!]
	\begin{tabular}{r l ||  c c c  }
	& & G1 & G2 & G3 \\
	\hline
	& {Virial quantities} & & & \\
	1.& log$_{10}$ (M$_{stars}$ [M$_{\odot}$] ) & 10.60 & 10.20 & 9.80 \\
	2.& $R_{200}$ [kpc] & 99.8 & 73.4 & 54.0 \\
	3.& $M_{200}$ [10$^{10}$ M$_{\odot}$]& 102.4 & 40.8 & 16.2 \\
	4.& $V_{200}$ [km.s$^{-1}$] & 210.1 & 154.6 & 113.7 \\
	\hline
	& {scalelengths} [kpc]& & & \\
	5.& $r_{stars}$ & 2.28 & 1.62 & 1.15 \\
	6.& $r_{gas}$  & 3.71 & 2.64 & 1.88 \\
	7.& $h_{stars}$  & 0.46 & 0.32 & 0.23 \\
	8.& $h_{gas}$  & 0.19 & 0.13 & 0.09 \\
	9.& $r_{bulge}$  & 0.46 & 0.32 & 0.23 \\
	10.& $r_{halo}$  & 19.95 & 14.68 & 10.80 \\
	11.& $r_{cut,stars}$  & 11.13 & 7.92 & 5.63 \\
	12.& $r_{cut,gas}$  & 11.13 & 1.94 & 1.38 \\
	13.& $h_{cut,stars}$  & 2.73 & 7.92 & 5.63 \\
	14.& $h_{cut,gas}$  & 0.56 & 0.40 & 0.28 \\
	15.& $r_{cut,bulge}$  & 2.28 & 1.62 & 1.151 \\
	16.& $r_{cut,halo}$  & 49.88 & 36.69 & 26.99 \\
	17.& $r_{metal}$  & 3.71 & 2.64 & 1.88 \\
	\hline
	& {Mass fractions} & & & \\
	18.& $f_g$ & 0.65 & 0.65 & 0.65 \\
	19.& $f_b$ & 0.10 & 0.10 & 0.10 \\
	20.& $m_d$ & 0.10 & 0.10 & 0.10 \\
	
	\hline
	& {Collision-less particles} [10$^6$] & & &\\
	21.& $N_{disk}$  & $2.00$ & $0.80$ & $0.32 $ \\
	22.& $N_{halo}$ & $2.00$ & $0.80$ & $0.32$ \\
	23.& $N_{bulge}$ & $0.22$ & $0.09$ & $0.04 $ \\
	
	\hline
	& {Various quantities} & & &\\
	24.& $Q_{min}$ & 1.5 & 1.5 & 1.5 \\
	25.& $c$ & $5$ & $5$ & $5$ \\
	26.& $Z_{core}$ & $0.705$ & $0.599$ & $0.479$ \\
	
	\end{tabular}
	\vspace{0.3cm}
	\caption{Physical properties of the three high-redshift disk models (G1,G2,G3). All the quantities based on the cosmology use $\Omega_{\Lambda}=0.7$ and $\Omega_m=0.3$ and z=2. \newline
	1. $M_{stars}$ is stellar mass. 
	2. Virial radius (radius at which the density of the halo reaches 200 times the critical density of the Universe).
	3. Cumulated mass at the virial radius.
	4. Circular velocity at the virial radius.
	5. Stellar disk scalelength.
	6. Gaseous disk scalelength.
	7. Stellar disk scaleheight.
	8. Gas disk scaleheight.
	9. Stellar bulge scalelength.
	10. Dark matter halo scalelength.
	11. Stellar disk radial cut.
	12. Stellar disk azimuthal cut.
	13. Gas disk radial cut.
	14. Gas disk azimuthal cut.
	15. Stellar bulge radial cut.
	16. Dark matter halo radial cut.
	17. Metallicity scalelength.
	18. Gas fraction.
	19. Stellar bulge mass fraction.
	20. Baryonic mass fraction: a mass fraction $m_d$ of $M_{200}$ mass is considered to be in a disk.
	21. Number of particles in stellar disk.
	22. Number of particles in dark matter halo.
	23. Number of particles in stellar bulge.
	24. Minimal value for the Toomre stability parameter in the initial conditions.
	25. Concentration parameter of the halo.
	26. Fraction of metals in the gas at the center of the galaxy, in units of solar metallicity.
	}
	\label{properties_simulations}
\end{table}

\subsection{Orbital parameters}
{The MIRAGE sample is designed to constrain the kinematical signatures induced by a galaxy merger on rotating gas-rich disks. To this purpose, we build a sample which explores probable disks orientations which are likely to produce a wide range of merger kinematical signatures.}
It has been statistically demonstrated using dark matter cosmological simulations that the spin vectors of the dark matter haloes are not correlated one to the other when considering two progenitors as Keplerian particles \citep{2006A&amp;A...445..403K}. We use this result to assume that no spin orientation configuration is statistically favored.
Our  galaxy models are placed on Keplerian orbits using $\theta_1$ the angle between the spin vector of the first galaxy and the orbital plane, $\theta_2$ the angle between the spin vector of the second galaxy and the orbital plane, and $\kappa$ the angle between the spin vector of the first galaxy and the second one (see Fig. \ref{simus_geometry}).
If these angles are uncorrelated, the normalized spin vectors are distributed uniformly over the surface of a sphere. 
Consequently, all the spin orientations are equally probable. 
{If one consider a random sampling of these disk orientations using a small finite solid angle, the spin vector being coplanar to the orbital plane produces the largest number of configurations. Therefore, we favor configurations where we have at least one spin vector in the orbital plane, i.e. $\theta_1=90^{\circ}$ in all the configurations. We specifically avoid configurations where both of the disks are in the orbital plane because they are highly unlikely and are subject to strong resonances not statistically relevant.
We assume that the fourth angle $\omega$ which orientates the first galaxy with respect to its line of node  \citep{1972ApJ...178..623T} might not degenerate the kinematics and the shape of the merger remnant since this parameter does not affect the total angular momentum of the system. Consequently, we arbitrarily choose to have the spin vector of the first galaxy always collinear to its Keplerian particle velocity vector.}
We define each orbit name by the concatenation of the angles $\theta_1$, $\theta_2$ and $\kappa$ (see Table \ref{orbits}).

\begin{figure*}
	\centering
	\includegraphics[width=18cm]{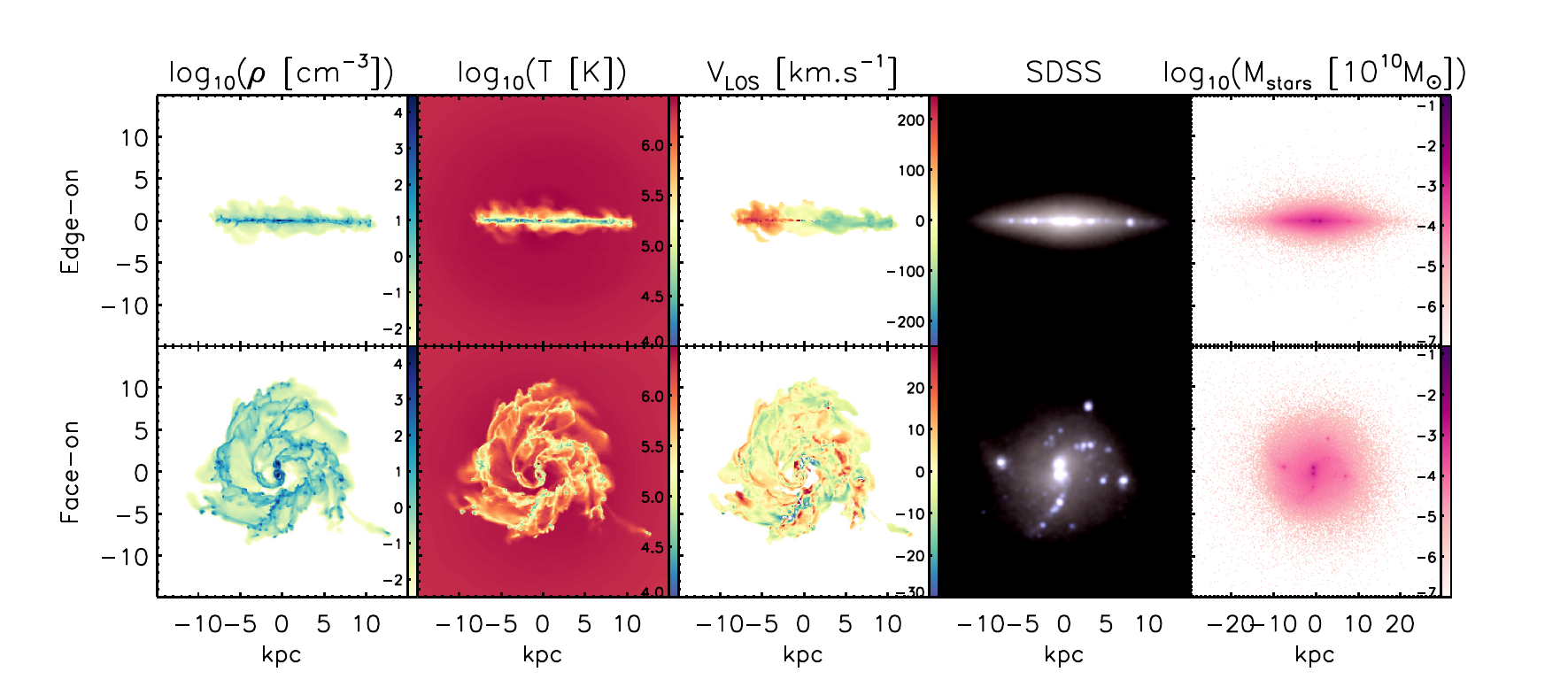}
	\caption{Maps for the G2 model after 400 Myr of evolution. From left to right: mass-weighted mean gas density, mass-weighted mean gas temperature, mass-weighted mean gas radial velocity, SDSS u/g/r mock observation built from the STARBURST99 model using stellar particles age and mass and assuming solar metallicity, stellar mass map. The upper line presents an edge-on view, while bottom line displays a face-on view.}
	\label{band}
\end{figure*}

\begin{figure*}
	\centering
	\includegraphics[width=14cm]{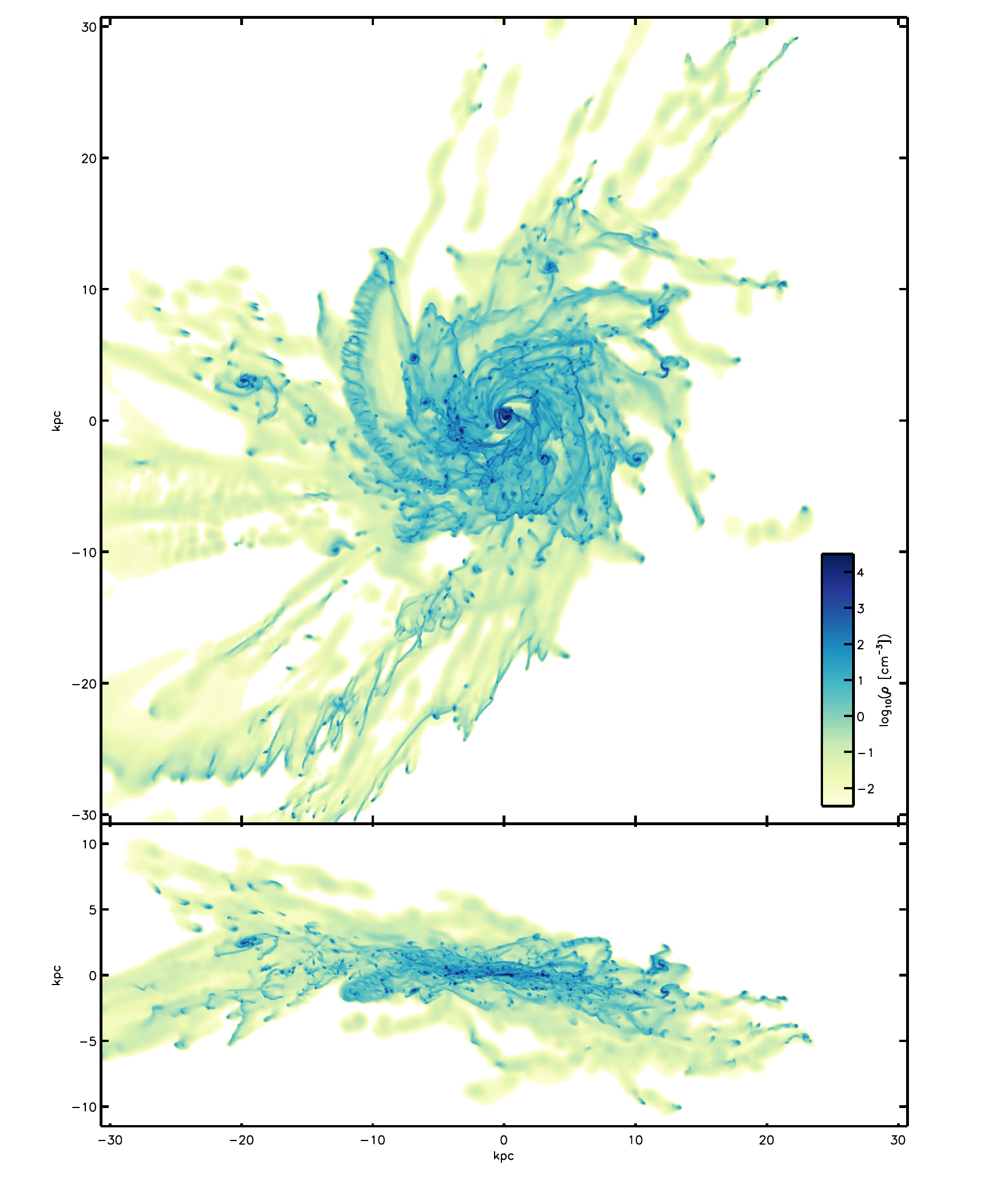}
	\caption{Face-on (top) and edge-on (bottom) mass-weighted average density maps of the gas for the G1\_G1\_90\_90\_0 simulation 280 Myr after the coalescence (i.e. 640 Myr of evolution after the initial conditions).}
	\label{dens_large}
\end{figure*}

The choice of studying a wide range of spin vector orientations is motivated by the requirement to detect  extreme signatures, and to bind the kinematical and morphological parameters of the merger remnants. However, we introduce a random angle $\delta$ when setting up the spin vector of our galaxies, picked using a uniform distribution introducing a $\pm5^{\circ}$ uncertainty. This method is implemented to prevent  alignment with the AMR grid which could produce spurious effects. The slight un-alignment also increases the numerical diffusion typical of grid codes, which in our case can help to relax our initial conditions. 

The pericenter distance, i.e. the distance between the two galactic centers at the time of the closest approach along the Keplerian trajectory is chosen to be $r_{peri}=r_{1,cut,gas}+r_{2,cut,gas}$, where $r_{1,cut,gas}$ and $r_{2,cut,gas}$ are respectively the cut radii of the first and the second galaxy. This parametrization ensures that  disks do not collide on the first pericentral passage, even if the disks are already fragmented at the pericentral time. This choice is supported by the argument that low pericentral distances are not statistically relevant because the collision cross section is directly proportional to the value of this parameter, i.e. low pericentral distance is less probable. By specifying one value for the specific orbital energy of the system, one can compute the eccentricities of the orbits. To better understand the effects of the interaction parameters on the kinematics of our merger remnants, we set this parameter to a fixed negative value :
\begin{equation}
	E^{*} = \frac{v_{ini}^2}{2} - \frac{G(m_1+m_2)}{r_{ini}} = -2.85 \times 10^4 \; \mathrm{km^2.s^{-2}}\;,
	\label{orb_energy}
\end{equation}
where $E^{*}$ is the specific orbital energy, $v_{ini}$ is the initial relative velocity of the galaxies, and $r_{ini}$ is the initial distance between the galaxies. This negative specific orbital energy means that all of our trajectories are elliptic ($e<1$). The parameters of the Keplerian orbits are listed in Table \ref{orbital_params}. We acknowledge that such low eccentricities might not be statistically relevant \citep{2006A&amp;A...445..403K}, but we save computational time. 

\begin{table}[h!]
	\begin{tabular}{l ||  c c c c c }
	orbit label & $\theta_1$ & & $\theta_2$ & & $\kappa$ \\
	\hline
	90\_90\_90 & $90^{\circ}$ &  & $90^{\circ}$ & & $90^{\circ}$ \\
	90\_90\_0 & $90^{\circ}$ & & $90^{\circ}$ & & $0^{\circ}$ \\
	90\_90\_180 & $90^{\circ}$ & & $90^{\circ}$ & & $180^{\circ}$ \\
	90\_ 0\_90 & $90^{\circ}$ & & $0^{\circ}$ & & $90^{\circ}$ \\
	\end{tabular}
	\vspace{0.3cm}
	\caption{Orbital angles describing the four orbits studied in this paper. We introduce a random deviation $|\delta|<5^{\circ}$ (not given in the table) in our merger setup in order to avoid over symmetry of our simulations. The orbit name is the concatenation of the angles $\theta_1$, $\theta_2$ and $\kappa$.}
	\label{orbits}
\end{table}  

\begin{table}[h!]
	\begin{tabular}{l ||  c c c c c }
	 & $r_{ini}$  & $v_{ini}$ & $r_{peri}$ & $e$ & $E$ \\
	 
	 & \footnotesize[kpc] & \footnotesize[km.s$^{-1}$] & \footnotesize[kpc] & &\footnotesize[10$^4$ kg.km$^2$.s$^{-2}$] \\
	\hline
	G1\_G1 & 68 & 237 & 21.8 & 0.67 & -63.1	\\
	G1\_G2 & 57 & 204 & 19.1 & 0.61 & -38.0	\\
	G1\_G3 & 52 & 183 & 16.8 & 0.59 & -18.2	\\
	G2\_G3 & 26 & 173 & 13.6 & 0.31 & -15.1	\\
	G2\_G2 & 36 & 167 & 15.5 & 0.42 & -25.1	\\
	\end{tabular}
	\vspace{0.3cm}
	\caption{Orbital parameters of the five configurations explored in the MIRAGE sample. These parameters are obtained using $E^{*}=-2.85\times$ 10$^4$ km$^2$.s$^{-2}$ and $t_{peri}$=250 Myr. $r_{ini}$ is the initial distance between the two galaxies, $v_{ini}$ is the initial relative velocity of the two galaxies, $r_{peri}$ is the pericenter distance, $e$ is the eccentricity of the orbits, and $E$ is the orbital energy of the system.}
	\label{orbital_params}
\end{table}

Finally, we define the initial distance between galaxies with a conservative expression through different merger masses, using the pericenter time, i.e. the time for the galaxies to reach the pericenter, with $t_{peri}$=250 Myr. The choice of the specific orbital energy is achieved in order to be able to set $t_{peri}$ to 250 Myr with elliptic orbits $e<1$. Because the dynamical times of all the models are close, this formulation ensures that the models relax synchronously before the start of the interaction (see section \ref{relax}). 
Our sample encompasses 20 merger configurations (four sets of orbital angles, five sets of orbital parameters due to different galaxy masses), to which we add the three isolated disk models in order to have a reference for secular evolution. We have excluded the G3\_G3 interaction to save computation time, since the relative resolution on the merger remnant is coarser than any other cases.

\subsection{Environment}
\label{environment}
{We aim at simulating the accretion from an idealized hot gaseous halo surrounding the galactic disks. To this purpose, we model the inter-galactic medium (IGM) by setting an initial minimum gas density $\rho_{IGM}=2.3\times 10^{-4}\mathrm{cm}^{-3}$ within the AMR box. The gas present in the IGM is initialized with no velocity, so that it collapses towards the central potential well at the free-fall velocity. After a dynamical time, the gas halo reaches a state close to a spherical hydrostatic equilibrium where the densest regions are allowed to cool down. The zero gradient condition imposed in the grid boundaries implies a continuous injection of pristine gas on the boundaries of the AMR box.}

\section{Evolution of global physical properties}
\label{properties}

Fig. \ref{band} shows the morphology of the gas and the stars after 400 Myr of evolution along two orthogonal line-of-sight (LOS) for the simulation G2. With the first LOS, we see the disk edge-on, while the second LOS provides a face-on view. For each LOS, the gas density, temperature, as well as the morphology of the stellar component through a rest-frame SDSS mock composite ($ugr$ bands) image are displayed. We use a pixel of 0.396", and we project our simulations to a luminous distance of 45 Mpc, which gives a pixel size of 0.12 kpc assuming WMAP9 cosmological parameters values. The physical quantities computed for the gas are all mass-weighted averages along the LOS. The stellar emission is computed using the STARBURST99 model \citep{1999ApJS..123....3L} given the age and the mass of each particle. Contrary to \cite{2013MNRAS.tmp..733H}, we choose to neglect the dust absorption in the building of the SDSS mock images to emphasize the stellar light distribution. Projections misaligned with the AMR grid are always difficult to build. To palliate this common issue, we use multiple convolutions with smoothing kernel sizes adapted to the cell sizes.

The projections in Fig. \ref{band} show a disk with clumps lying in a turbulent medium. The most massive clumps reach masses of $\sim10^9M_{\odot}$ \citep{2013arXiv1307.7136B}. We observe a gaseous disk thickened by stellar feedback. The edge-on velocity field shows nevertheless a clear ordered rotation. The clumps concentrate most of the stellar emission due to young stars, since they host most of the star formation. {Fig. \ref{dens_large} further emphasizes this highly complex behavior of the gas with a substantial turbulence and disk instabilities by showing the mass-weighted average density of one of the the most massive merger simulation among our sample (G1\_G1\_90\_90\_0). We observe star-forming clumps wandering in a very turbulent ISM where the spiral structures are continuously destroyed by the cooling induced fragmentation and the thermal energy injection from stellar feedback. The edge-on view displays a disk thickened by the tidal torque induced by the recent merger.}
In the appendices of this paper (appendix \ref{append}), projections similar to Fig. \ref{band} are given for three simulations of the MIRAGE sample, covering the evolution up to 800 Myr, displayed on 11 time steps. The whole MIRAGE sample maps, containing 23 figures, are available as online material.

\subsection{Initial conditions relaxation}
\label{relax}
The relaxation of the disk plays a {fundamental} role at the beginning of the simulation. The low halo concentration when compared to lower-redshift, combined with a high gas fraction drives the gas disk towards an unstable state with $Q<1$, despite the fact that we start our simulation with the requirement $Q>1.5$ everywhere in the stellar disks. {The high cooling rates of the gas in the initial disk allow a very fast dissipation of internal energy.} To prevent a too rapid uncontrollable relaxation, we start our simulations with a maximum resolution of 59 pc ($\ell_{max}=12$) and with a temperature floor for the gas of $T=10^4 \mathrm K$ (see Table \ref{relax_strategy}). To establish smoothly the turbulence afterwards, we increase progressively the resolution every 25 Myr starting from 85 Myr, until we reach a maximum resolution of 7.3 pc, and a temperature floor for the gas of $300$ K. This allows the disks to form quickly spiral features supported by the thermal floor during the first time steps. Once the resolution is increased, Jeans instabilities arise and give birth to clumps of a few $10^8$ solar masses, which can quickly merge to form more massive ones. We observe a rapid contraction of the disks, reducing their radial size by $\sim20\%$ during the first 80 Myr, due to the dissipation of energy by the gas component. {This ad-hoc relaxation strategy insures to dissipate gradually the internal energy of the gas disk through cooling over 130 Myr, and also helps us to save computational resources.}
The refinement down to the level $\ell=14$ at $t=105$ Myr allows to reach densities $\rho>\rho_0$, enabling the formation of stars, and all the associated feedback of newly formed stars.

\begin{table}[h!]
	\begin{tabular}{l ||  c c c c c  }
	t [Myr] & $\ell_{max}$ & & $\Delta x$ [pc] & & $T_{floor}$ [K]\\
	\hline
	$[0,80]$ & 12 & & 58.6 & & $10^4$ \\
	$[80,105]$ & 13 & & 29.3 & & $10^3$ \\
	$[105,130]$ & 14 & & 14.6 & & $500$ \\
	$[130,800]$ & 15 & & 7.3 & & $300$ \\
	\end{tabular}
	\vspace{0.3cm}
	\caption{Refinement strategy of the high-redshift disks. At a third of the dynamical time of our models (85 Myr), we start to increase the maximum resolution and we lower the temperature floor allowing the gas to cool and dissipate internal energy.}
	\label{relax_strategy}
\end{table}  

\begin{figure*}[htbp]
	\centering
	\includegraphics[width=16cm]{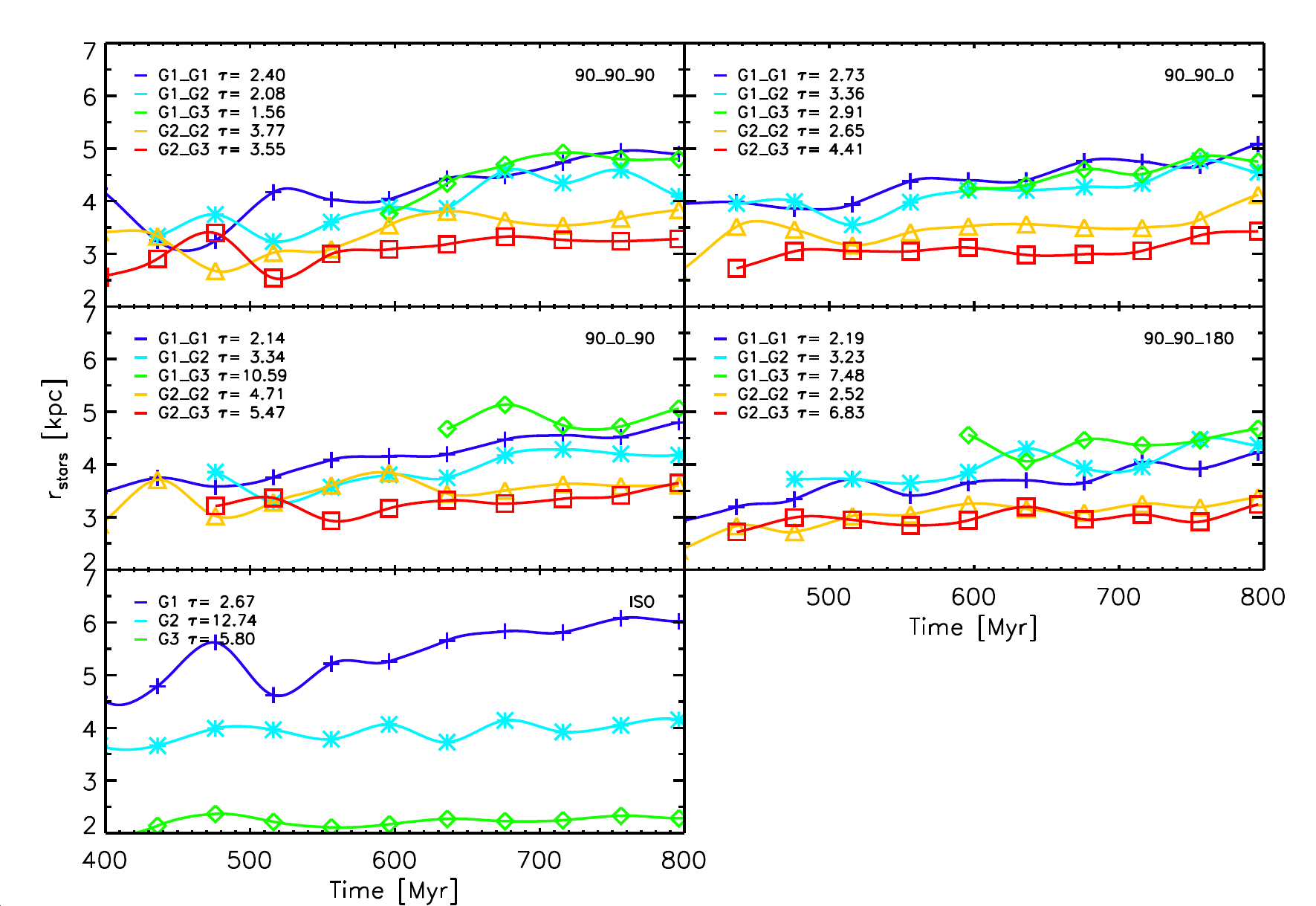}
	\caption{Evolution of the stars disk scale-length in the MIRAGE sample. Each panel traces the evolution of the scalelength for a given orbital configuration, allowing a comparison between mass ratios for a given set of disk orientations at a given specific orbital energy. The measurements are performed each 40 Myr, starting at the time of the core coalescence (400 Myr for the fastest mergers), and each curve linking these measurements is the result of a cubic interpolation to increase the clarity of the plot. The colored lines and different symbols indicate the mass ratio of the progenitors (given by G$i$\_G$j$, see Table \ref{orbital_params}); the label at the top right of each panel indicates the initial orientation of the disks (given by $\theta_1 \_ \theta_2 \_ \kappa$, see Table \ref{orbits}). The lower left panel is dedicated to isolated simulations. For each simulation, we indicate the growth time $\tau$ expressed in Gyr, which is the time needed for the disk/remnant to double its size starting from the closest measurement to 400 Myr.}
	\label{time_dscale}
\end{figure*}

\subsection{Gas accretion from hot halo}
\label{accretion_section}

As mentioned in section \ref{environment}, the AMR box is continuously replenished with metal-free gas. The very low density component is constrained by a gamma-polytrope which ensures the formation of a hot stabilized halo. The central part of the halo reaches densities above $10^{-3}$ cm$^{-3}$, where the pressure support from the gamma-polytrope is ended. Thanks to metal lines cooling, the central part of the gaseous halo can cool down and condense on the galactic disk. We measure the accretion rates in a spherical shell of  20 kpc (typical value of the halo scalelength in the most massive galaxy model), by detecting the metal poor gas (Z<10$^{-3}$) able to enter the sphere within a time step of 5 Myr. In Table \ref{mean_accretion} we display the mean values of these accretion rates as well as the mean SFR for the different masses configurations of the sample. We compare these values to theoretical predictions from the baryonic growth rate formula found in \cite{2009Natur.457..451D}. The theoretical values are obtained using total halo masses (without considering radial cuts), and we assume that only two thirds of this accretion rate can be associated to smooth gas accretion, the remaining third being associated to mergers as observed in \cite{2009Natur.457..451D}.

\begin{table}[h!]
	\begin{tabular}{l || r r r r}
	
	& $\langle \dot{M}_{gas} \rangle$   & $\langle$SFR$\rangle$ & $\dot{M}_{th}(z=2)$ & $\dot{M}_{th}(z=1.5)$\\
	& [M$_{\odot}$.yr$^{-1}$] & [M$_{\odot}$.yr$^{-1}$] & [M$_{\odot}$.yr$^{-1}$] & [M$_{\odot}$.yr$^{-1}$]\\
	\hline
	G1 		 & 	13.4 		& 17.5 	& 31.8	& 21.1	\\
	G2 		 & 	2.6   		& 7.7		& 11.0	& 7.3		\\
	G3 		 & 	0.8   		& 3.7		& 3.8		& 2.5		\\
	G1\_G1 	 & 	41.4 		& 31.5 	& 63.6	& 42.2	\\
	G1\_G2 	 & 	20.8 		& 23.2 	& 42.8	& 28.4	\\
	G1\_G3 	 & 	17.4 		& 20.2 	& 35.6	& 23.6	\\
	G2\_G2 	 & 	8.0   		& 13.9 	& 22.0	& 14.6	\\
	G2\_G3 	 & 	6.0   		& 10.7 	& 14.8	& 9.9		\\

	\end{tabular}
	\vspace{0.3cm}
	\caption{{Comparison of the mean star formation and accretion rates measured in the MIRAGE sample.} $\langle \dot{M}_{gas} \rangle$: average accretion rate of inflowing pristine gas (Z<10$^{-3}$) for the isolated disks and the mergers measured in a spherical shell with a radius of 20 kpc. $\langle$SFR$\rangle$: average SFR. $\dot{M}_{th}(z)$: Theoretical prediction of the gas accretion rate as function of redshift and  halo mass. All the averages are computed in the interval [100,800] Myr.}
	\label{mean_accretion}
\end{table}

Furthermore, at z=2, \cite{2009MNRAS.397L..64A} find an accretion rate of hot gas for a galaxy with a baryonic mass $M_b\sim10^{11}$ $M_{\odot}$ in good agreement with our model having a close mass (namely the G1 model). They also show that cold gas accretion which prevails at $z\gtrsim2$ becomes dominated by hot gas accretion at lower redshifts, which makes our implementation in agreement with this statement. This scenario is also supported by recent works using moving-mesh code which find a substantially lower cold gas accretion rate than in comparable SPH simulations \citep{2013MNRAS.429.3353N}. The gas accretion rate in the MIRAGE sample slightly increases with time (see section \ref{environment}), implying that simulating more than 1 Gyr of evolution would lead to unrealistic high accretion rates. The mean accretion rates measured over 800 Myr in our simulations remain consistent with theory and cosmological simulations (see Table \ref{mean_accretion}).

\normalfont
\subsection{Mass-size evolution}
\label{mass_size}

\begin{figure*}[htbp]
	\centering
	\includegraphics[width=18cm]{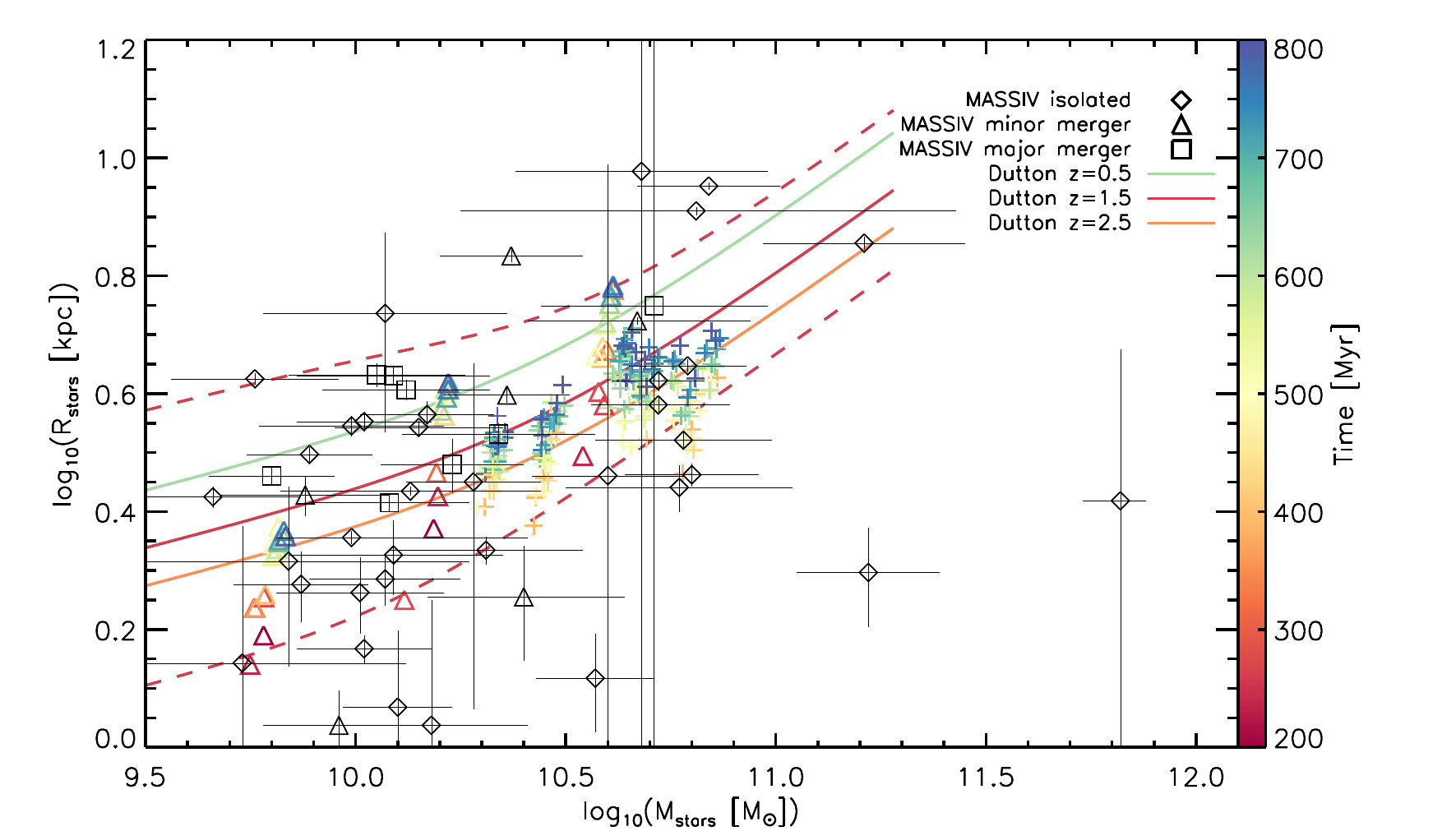}
	\caption{Stellar mass as a function of stellar scalelength. The symbols ``+'' and ``$\Delta$'' show respectively the MIRAGE galaxy mergers and isolated disks. The color is encoding the time evolution since the initial conditions. Black symbols display the MASSIV data, according to measurements found in \cite{2012A&amp;A...546A.118V} and \cite{2012A&amp;A...539A..92E}. The stellar mass-size relation derived in \cite{2011MNRAS.410.1660D} and shifted to $z=1.5$ is over-plotted with the red solid curve. The dotted curves show the dispersion computed for $z=1.5$ from the relation derived in \cite{2011MNRAS.410.1660D}. We also display the mass-size relation for $z=0.5$ (green line) and $z=2.5$ (orange line) to emphasize the redshift evolution of the relation.}
	\label{mstar_rstar}
\end{figure*}

To follow the evolution of the mass-size relation in our simulation sample, we proceed to a centering and spin alignment with the z-axis of the AMR cartesian grid. As the stars age, they experience a progressive gravitational heating which redistributes the oldest stars into a diffuse halo component with a smoother gravitational potential. The center of the each simulation is therefore found using the peak of the mass-weighted histogram of the positions of the oldest stars, i.e. the stars only present in the initial conditions. This peak in the distribution of the old stars is hereafter associated to the center of the bulge. We recover the galactic disk orientation using the spin vector of the stars younger than 50 Myr for a given snapshot. They tend to be still located within the gas disk which can not be used to perform such computation because of the turbulence and the outflows carrying consequent momentum. Once the orientation of the galactic disk is correctly recovered, we compute the stellar surface density profile. {The distribution is decomposed into a bulge and a disk by performing a linear regression on the surface density profile in the interval [$r_{cut,bulge}$,$r_{cut,stars}$] to extract the disk profile (see Table \ref{properties_simulations}). To each surface density measurement we associate a relative error proportional to the square root of the number of particles found within the radial bin.}

The evolution of the disk scalelengths is displayed in Fig. \ref{time_dscale}. It testifies that both mergers and isolated disks can produce an inside-out growth \citep{2006MNRAS.366..899N} regardless the orbital configuration, despite the proven ability of gas-rich mergers to produce compact systems \citep{2011ApJ...730....4B}.
For each simulation, we estimate the growth time, which we define as the time needed for the stellar disk to double its size measured right after the coalescence (or at t=400 Myr for the isolated systems). A mean growth time of 3.9 Gyr is measured for the mergers, with the fastest systems reaching growth time close to 2 Gyr. It appears that the less massive systems, therefore the less clumpy, are less efficient to drive stellar disk growth.
This inside-out growth is taking place in an idealized framework, although the galaxies are accreting gas from the halo at a rate comparable to cosmological simulations (see Section \ref{accretion_section}). This continuous gas accretion fuels secular evolution processes that are able to drive such growth by performing a mass redistribution. Consequently, our results suggest that other mechanisms than late infall of cold gas from the cosmic web \citep{2011MNRAS.418.2493P} may alternatively build up high-redshift disks inside-out.

The stellar mass-size relation for the MIRAGE sample is shown in Fig. \ref{mstar_rstar}. We plot the mass-size relation found in \cite{2011MNRAS.410.1660D} and shifted at different redshifts.  
Our choice of stellar sizes in the initial conditions makes the simulations of the MIRAGE sample lie in the dispersion range computed for the $z=1.5$ mass-size relation. Nonetheless, the size evolution is fast in the MIRAGE sample, but one can expect this rapid growth to stop once the clumpy phase ends. Indeed, the size growth is linked to the gas-rich clump interaction which is able to redistribute significant amount of stellar mass towards the outskirts of the disk.
We over-plot on the simulations data the values for the MASSIV sample; the error bars show the $1\sigma$ standard deviation computed using the errors on the stellar mass and size \citep{2012A&amp;A...546A.118V}. {We use the classification of \cite{2013A&amp;A...553A..78L} to differentiate isolated galaxies from minor and major mergers on the plot using different symbols. We observe that the majority of galaxies classified as major merger lie above or close to the $z=1.5$ mass-size relation, which is straightforward once considered that the size measurement is done on a extended system where the two disks are not yet well mixed. This gives credit for the major merger classification performed by \cite{2013A&amp;A...553A..78L}.  Overall, the bulk of the MASSIV sample ranges within the dispersion fork of the $z=1.5$ relation, which makes our simulations consistent with observations. A fraction of the isolated and minor merger systems are more than $1\sigma$ below the $z=1.5$ relation, suggesting a population of compact galaxies.}

\begin{figure*}[htbp]
	\centering
	\includegraphics[width=16cm]{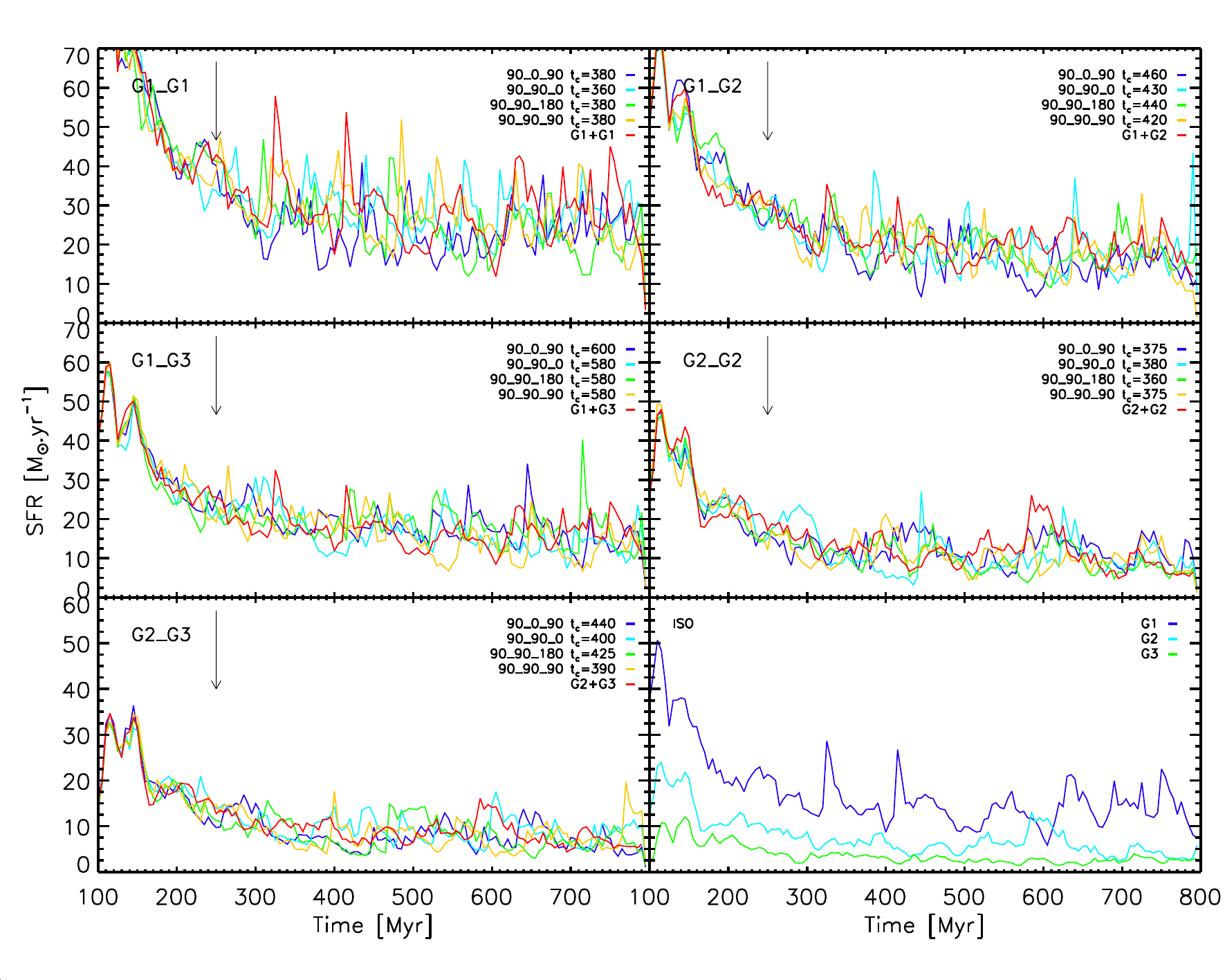}
	\caption{Star formation histories for each simulation of the MIRAGE sample. Each panel explores disk orientations for fixed masses respectively given by $\theta_1 \_ \theta_2 \_ \kappa$ and G$i$\_G$j$ (written on the top right of each panel, see Table \ref{orbits} and \ref{orbital_params}). The last panel shows the SFR of the isolated disk simulations. The curves begin at 100 Myr (see section \ref{relax}). In order to compare the SFR of merging disks with the SFR of isolated disk per mass unit, the SFR of isolated disks (red dotted lines) have been superimposed to the SFR of merging disks. The black arrow in the merger panels shows the pericentral time $t_{peri}$ equal to 250 Myr in all the merger simulations. For each galaxy merger, we also display the time of the coalescence of the galactic cores $t_{c}$ visually determined.}
	\label{time_sfr}
\end{figure*}

\subsection{Star formation}

Fig. \ref{time_sfr} presents the star formation histories of the MIRAGE sample for the different masses and merger orbital configurations. {For each simulation, we indicate in the legend the coalescence time $t_c$ expressed in Myr, as well as the pericentral time $t_{peri}$ shown by a black arrow.} The star formation histories exhibit a stochastic behavior due to the clump interactions and the cycling energy injection by stellar feedback maintaining the gas turbulence. The mean ratio of the SFR dispersion over the average SFR ($\sigma_{\mathrm{SFR}}/\langle \mathrm{SFR} \rangle$) for the whole MIRAGE sample is roughly equal to 30\%. Quite surprisingly at first glance, we do not observe any SFR enhancement due to the galaxy merger. Neither orientations, nor mass configurations appears to produce enhanced SFR.

\noindent {Fig. \ref{relative_sfr_histo} shows the histogram of the normalized quantity in the interval $t_\mathrm{c}\pm100$ Myr: $$(\mathrm{SFR}(t)-\mathrm{SFR}_{\mathrm{iso}}(t))/\langle \mathrm{SFR}_{\mathrm{iso}}\rangle$$ 
with $\mathrm{SFR}_{\mathrm{iso}}$ the summed SFR of the fiducial simulations evolved in isolation, and $\langle\mathrm{SFR}_{\mathrm{iso}}\rangle$ the mean value of $\mathrm{SFR}_{\mathrm{iso}}$. For each histogram, we display in the legend the value of barycenter of the distribution $\beta$, which allows to estimate how much the interaction enhances the star formation in the time interval defined previously.}
We observe no trend of SFR enhancement due to merger ($\beta \leq 0$), even if in the case G1\_G3 two of the merger produces somewhat more stars than the summed fiducial isolated models ($0.1$$\leq$$\beta$$\leq$$0.2$). However, this value is too low to be considered as a starburst. Generally, the mergers are even less effective at producing stars compared to isolated simulations.
This result contradicts other works \citep[e.g.][]{2011ApJ...730....4B, 2010ApJ...720L.149T, 2008MNRAS.384..386C, 2013MNRAS.434.1028P}. As this paper does not intend to perform a full study of the starburst efficiency in high-redshift galaxies, we list subsequently and briefly discuss the possible reasons for the suppression of starburst in our simulation sample:

\begin{figure*}[htbp]
	\centering
	\includegraphics[width=16cm]{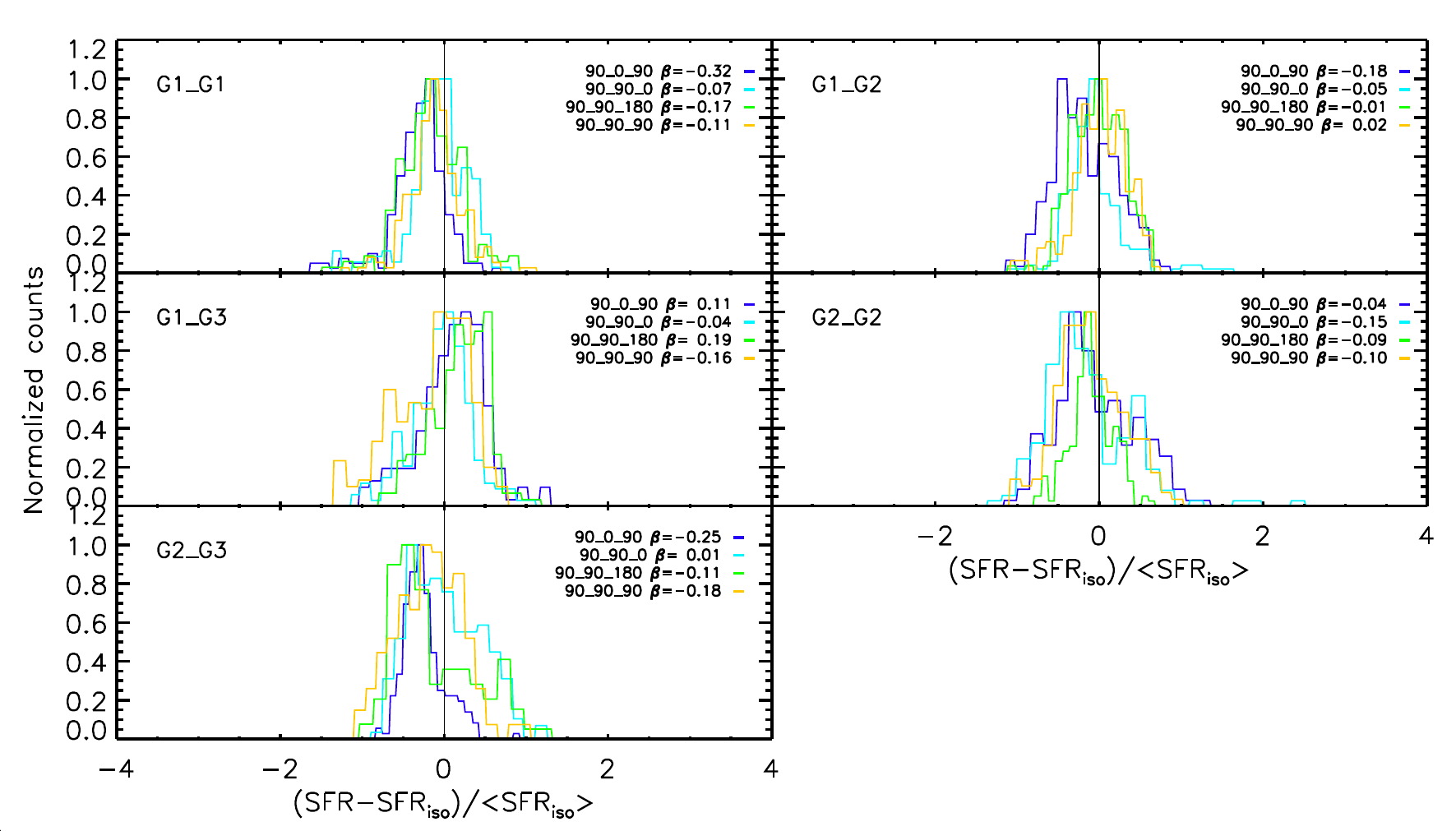}
	\caption{Histogram of the difference between the merger SFR and the cumulative isolated SFR ($\mathrm{SFR}-\mathrm{SFR}_{iso}$) computed between 100 Myr before and 100 Myr after the galaxies coalescence. Each panel explores disk orientations for fixed masses respectively given by $\theta_1 \_ \theta_2 \_ \kappa$ and G$i$\_G$j$ (written on the top left of each panel, see Table \ref{orbits} and \ref{orbital_params}). From these histograms, we can interpret how much time a merger spend with a higher or lower SFR during this crucial period. The quantity $\beta$ in the legend is the barycenter of the histogram, which measures the shift in star formation induced by the merger compared to secular evolution.}
	\label{relative_sfr_histo}
\end{figure*}

\begin{itemize}

\item Our choice of elliptic Keplerian trajectories might impact the star formation efficiencies of our merger simulations. However, many works demonstrated that the starburst efficiency of equal mass galaxy mergers is  insensitive to the orbits, the disk orientations, and the physical properties of these galaxies \citep[e.g.][]{1996ApJ...464..641M, 2000MNRAS.312..859S, 2004ApJ...607L..87C}. {Consequently, the initial configuration of the major merger simulations G1\_G1 and G2\_G2 should not be considered as responsible for the absence of starburst.} However, longer interactions would lead to coalescence of more concentrated systems because of the clumps migration, which could enhance a nuclear starburst. The fact that higher mass ratios simulations (G1\_G2, G1\_G3, G2\_G3) that explores more elongated orbits (see Table \ref{orbital_params}) do not  exhibit star formation enhancement might suggests that the orbits and the disk orientations are overall not to blame for this lack of star formation overactivity. 

\item As highlighted by \cite{2011MNRAS.415.3750M}, the hot gaseous haloes implied in a galaxy merger are likely to be heated by shocks, together with an acquisition of specific angular momentum increasing the centrifugal barrier. Both of these processes can push towards a lower starburst efficiency because isolated disks are more effective to accrete gas from the hot halo.

\item A complex treatment of the ISM favors the production of hot gas, which systematically lowers SF, as \cite{2006MNRAS.373.1013C} point out. The simulations performed in \cite{2011ApJ...730....4B} constitute a good dataset for direct comparison, due to the initial conditions definitions very close to our G1 model. The presence of starbursts in such comparable simulations when the gas obeys to a 1D equation-of-state suggests a change in the gas response to a galactic interaction.

\item \cite{2010ApJ...720L.149T} demonstrated that the starburst in a low-redshift major merger is mostly driven by the enhancement of gas turbulence and fragmentation as long as the numerical resolution allows to resolve it. It may be more difficult to increase this turbulence and fragmentation at high-redshift because both are already high in our isolated gas-rich disks. The isolated disks simulations are indeed able to maintain this high level of turbulence and fragmentation thanks to the continuous gas refilling by the hot halo accretion and an efficient stellar feedback. This scenario would suggest that star formation can saturate and prevent starbursts in galaxy mergers of very turbulent and clumpy gas-rich disks.

\item High gas fractions (>50\%) are maintained throughout the duration of the mergers. These high gas fractions may prevent the formation of a stellar bar in the remnant, which would drag large amount of gas toward the nucleus to fuel a starburst \citep{2009ApJ...691.1168H}. The large fraction of cold fragmented gas prevents the formation of a bar in the stellar component. Additionally, the stellar feedback removes gas from the star-forming regions continuously, and may also acts against the formation of a large stellar disk by lowering the SFR \cite{2011MNRAS.415.3750M}.

\item The feedback model adopted in this study might not be sufficiently energetic to succeed in ejecting important quantities of gas on very large scales especially because of the isotropic hot gas accretion which systematically curbs the outflowing material. The adopted feedback model may be efficient enough to saturate the star formation during the pre-merger regime, but is not strong enough to deplete the disk from large quantities of gas, which would then be re-accreted later feeding a star formation burst.

\end{itemize}

Numerous processes can explain the starburst removal in very gas-rich clumpy and turbulent galaxy mergers. The star formation histories of the MIRAGE sample remain difficult to interpret without a complete study in a full cosmological environment to weight each configuration according to its occurrence probability. Generally, the link between mergers and starburst may be more fuzzy at high-redshift than at lower redshift. 

\normalfont

\begin{figure*}[htbp]
	\centering
	\includegraphics[width=16cm]{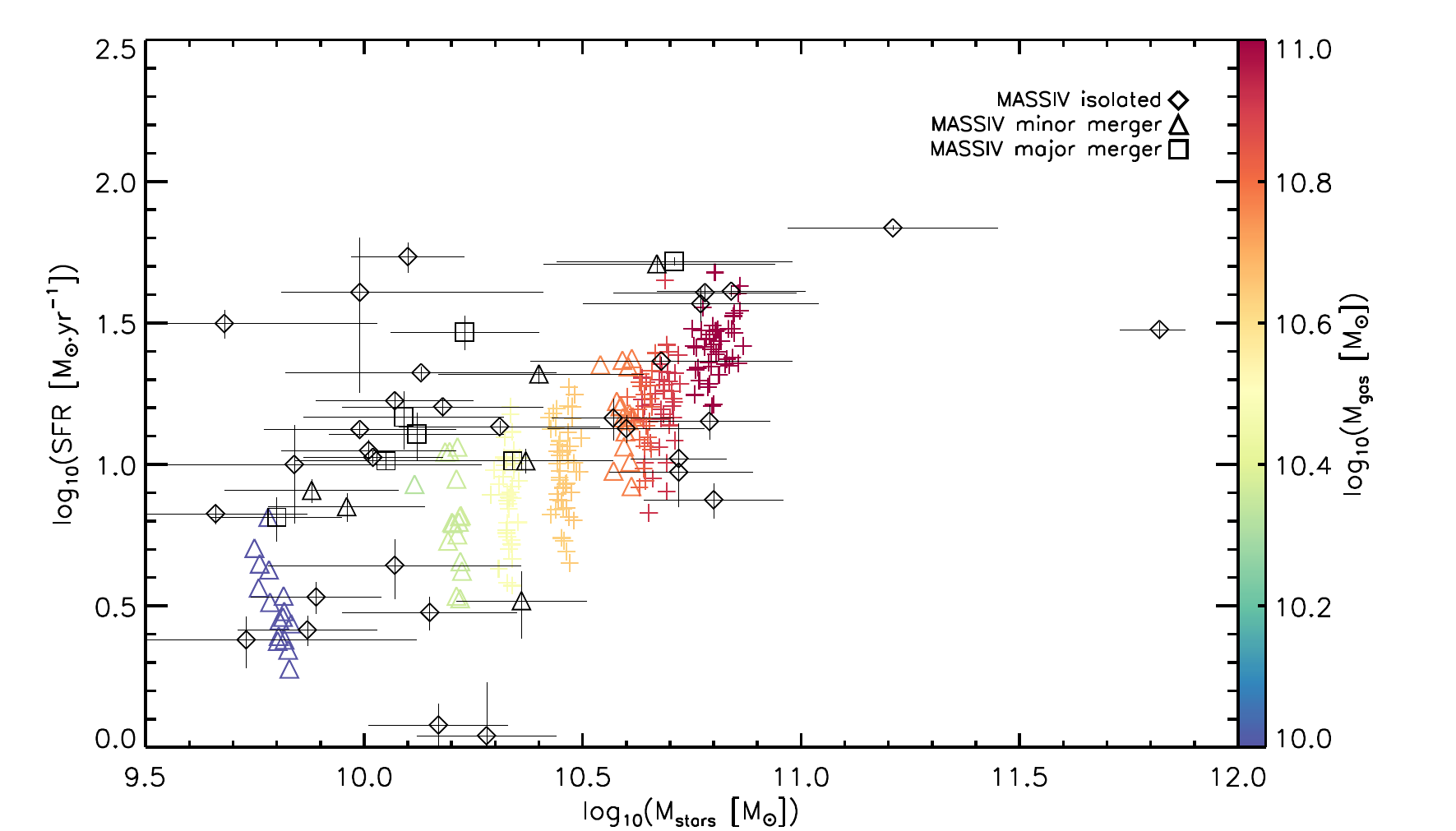}
	\caption{Star formation rate as a function of stellar mass measured between the coalescence and 800 Myr for the merger simulations, and between 200 and 800 Myr for the isolated simulations. Black symbols show MASSIV data for which the SFR is estimated from the $\mathrm{H\alpha}$ integrated luminosity, and the stellar masses measured within the optical radius $r_{opt}=3.2 \times r_{stars}$. Each colored symbol shows a snapshot of the MIRAGE mergers and isolated disks simulations, respectively plotted with ``+'' and ``$\Delta$''. The color encodes the gas mass of the disks and remnants measured within the gas optical radius.}
	\label{mstar_sfr}
\end{figure*}

Fig. \ref{mstar_sfr} displays the SFR as a function of the stellar mass. We compare the MASSIV ``first-epoch'' data with the MIRAGE sample, for which the SFR has been estimated from the integrated $\mathrm{H\alpha}$ luminosity, and stellar mass within the optical radius. The MASSIV error bars have been computed {using the errors on the $\mathrm{H\alpha}$ flux measurement found in \cite{2012A&amp;A...539A..93Q}}. The scatter observed for a given simulation stellar and gas mass underlines the stochastic nature of the star formation in gas-rich clumpy disks. This scatter is nevertheless still lower than the one observed in the MASSIV data which encompasses much various gas fractions.

Fig. \ref{ks_diagram} displays the position of the MIRAGE sample on the KS diagram between 200 Myr and 800 Myr for the isolated disks, and between the coalescence and 800 Myr for the merger simulations. 
We computed the gas surface density $\Sigma_{\mathrm{gas}}$ and the star formation surface density $\Sigma_{\mathrm{SFR}}$ quantities within $r_{stars}$, the stellar disk scalelength estimated with the method described in section \ref{mass_size}. We also rejected all the gas in the IGM by considering only cells with densities greater than $\rho=2\times10^{-3}$ cm$^{-3}$ which typically corresponds to the frontier between ISM and IGM in all of the simulations. The quantities $\Sigma_{\mathrm{gas}}$ and $\Sigma_{\mathrm{SFR}}$ are measured on face-on projections, after having centered our referential on the peak of the old stars probability distribution function, and aligned the spin of the young stars disks with our LOS. We note that our sample lies close to the relation found in \cite{2010ApJ...714L.118D}, with a slight shift towards lower star formation efficiencies (within the $1\sigma$ dispersion) which can be attributed to the shutdown of star formation at high gas temperatures. The MIRAGE sample does not show any bi-modality, as expected from the star formation histories displayed in Fig. \ref{time_sfr}. One should take into account that, by construction, our sample does not provide the statistical cosmological weight of a volume limited sample of the $1<z<2$ galaxy population.
{We also over-plot the position of the MASSIV sample on this diagram for comparative purposes. The MASSIV error bars on the quantity $\Sigma_{\mathrm{SFR}}$ are computed using again the uncertainties on the $\mathrm{H\alpha}$ flux. We also take into account an error proportional to the spatial sampling of the SINFONI data which we propagate to the measurement of the radius of the ionized emission region. We do not have a strong observational constraints for the amount of gas in the MASSIV galaxies. Nevertheless, we mark out the gas mass for each galaxy assuming a mean  gas fraction $f_g=45\%$, which is the mean value obtained on the dynamical/stellar mass diagram of the MASSIV sample \citep{2012A&amp;A...546A.118V}. Using the relation $M_{gas}=f_gM_{\mathrm{stars}}/(1-f_g)$, we can over-plot the MASSIV data on the KS diagram. We compute the errors bars of the $\Sigma_{\mathrm{gas}}$ quantity assuming a minimal gas fraction of $f_{g,min}=25\%$ and $f_{g,max}=65\%$, a range where we can expect the MASSIV sample to lie. We then propagate the errors on the stellar mass using $f_{g,min}$ and $f_{g,max}$. {Therefore, the global distribution of the MASSIV data on the KS relation is close to the ``normal'' regime of star formation, considering our assumptions on the gas fractions.} Our merger simulations match the area covered by both the isolated and merging galaxies of the MASSIV sample on the KS diagram.

\section{Summary and prospects}
\label{summary}

In this paper, we introduce a new sample of idealized AMR simulations of high-redshift ($1<z<2$) mergers and isolated disks referred to as MIRAGE (Merging and Isolated high-Redshift AMR Galaxies). {The sample is originally designed to study the impact of galaxy merger on the gas kinematics in a clumpy turbulent medium.} We focus this paper on the presentation of the methods used to build the MIRAGE sample and on the first results obtained on evolution of the masses, sizes, and star formation rates.

\noindent The key points of the goals and methods used in this paper can be summarized as follows:
\begin{itemize}
\item {We present the MIRAGE sample, a series of mergers and isolated simulations using the AMR technique in an idealized framework which aims at comparing extreme signatures in terms of gas kinematics. {The MIRAGE sample initial conditions probe four disk orientations (with $\kappa$ ranging from 0$^\circ$ to 180$^\circ$), five total baryonic merger masses (ranging from 4.9 to 17.5 $\times10^{10}\mathrm{M}_{\odot}$) and three galaxy mass ratios (1:1,1:2.5,1:6.3) among 20 merger simulations designed from three disk models (with baryonic masses of 1.4, 3.5, and 8.8 $\times10^{10}\mathrm{M}_{\odot}$).} The case of low gas fractions has been extensively studied in the literature, so we choose here to study only gas-rich galaxies ($f_g\sim60\%$) to study the impact of the presence of giant star-forming clumps in merging turbulent disks.}
 
\item We introduce DICE, a new public code designed to build idealized initial conditions. The initialization method is similar to what has been done in \cite{2005MNRAS.361..776S}. The use of MCMC algorithm to build a statistical distribution requiring only the 3D-density function as input allows us to consider in future developments the building of components with more complex density functions compared to the canonical ones used in this paper.
 
\item We use a new implementation of stellar feedback from the young, massive part of the IMF  \citep{2013MNRAS.tmp.2414R}, coupled to a supernova feedback with non-thermal processes modeled using a cooling switch \citep{2013MNRAS.429.3068T}. {The new physically-motivated implementation of young stars feedback allows us to track the formation of Strömgren spheres where the energy from the massive young stars is deposited, allowing future comparisons with simulations using feedback recipes parametrized with wind mass-loading factors.}
 
\end{itemize}

\noindent The key results of this paper can be summarized as follows:
\begin{itemize}
\item {Star formation in disks} --
We find that the star formation history of isolated disk galaxies is strongly fluctuating over the time for each simulation, {with a SFR dispersion close to 30\% around its mean value.} This star formation proceeds mostly in giant clumps of gas and stars and naturally gets a stochastic behavior. The small star formation bursts may account for the intrinsic scatter of the ``Main Sequence'' of star forming galaxies at $z=1-2$ \citep{2010ApJ...714L.118D}.

\item {Star formation in mergers} --
The minor and major gas-rich mergers of our sample do not induce major bursts of star formation significantly larger than the intrinsic fluctuations of the star formation activity. {The mechanisms for triggering active starburst at high-redshift could be different from the ones at low-redshift due to large differences in the amount of gas available for accretion in the circum-galactic medium that lies around stellar disks. This suggests that a complex modeling of the gas capturing a high level of fragmentation and turbulence maintained by stellar feedback and gas accretion may offer a mechanism of saturation for the star formation activity in high-z galaxies. 
The remarkable homogeneity of the observed specific SFR in high-redshift galaxies \citep{2007A&amp;A...468...33E, 2011A&amp;A...533A.119E,2012ApJ...745..182N} coupled to the prediction of a high occurrence of minor mergers in this redshift and mass range \citep{2009Natur.457..451D, 2009ApJ...694..396B} may support the scarcity of star formation bursts triggered by very gas-rich mergers.}

\item {Global star formation scaling laws} --
Overall, our sample of disks and mergers is compatible with the evolution of the mass-SFR relation observed for a complete sample of star-forming galaxies in the same mass and redshift range (namely the sample MASSIV, \citealt{2012A&amp;A...539A..91C}), and independently of the assumed disk and merger fraction in the sample. On a Kennicutt-Schmidt diagnostic, the majority of mergers are close to the ``normal'' regime of disk-like star formation as defined by \cite{2010ApJ...714L.118D}, \cite{2010MNRAS.407.2091G}, with a slight deviation towards lower star formation efficiencies.

\item {Size evolution} --
A stellar mass-size relation in accordance with \cite{2011MNRAS.410.1660D} is obtained in our models, and the evolution with redshift of this relation is also reproduced. In particular, inside-out growth can be obtained as a natural outcome of internal dynamical processes redistributing angular momentum mostly through clumps interactions: these processes can naturally make disks become larger over time, for any given stellar mass, even if mergers are expected to produce more compact systems. {Our simulations include only infall of low-angular momentum material through hot gas accretion, suggesting that the radial {inside-out growth} at the observed rate might not need to be achieved through a cold mode in the context of our idealized modeling.}

\end{itemize}

 \begin{figure*}[htbp]
	\centering
	\includegraphics[width=16cm]{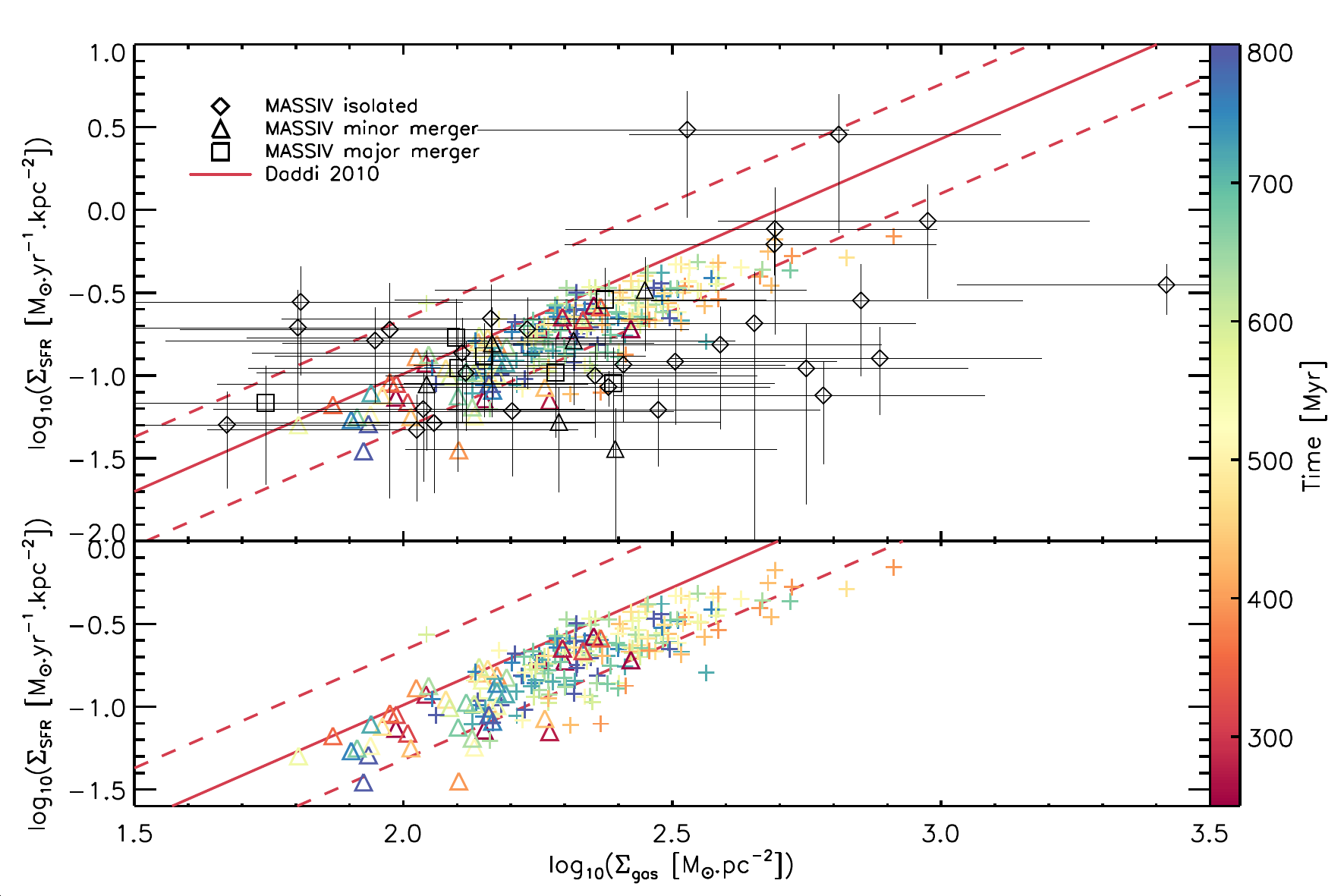}
	\caption{Kennicutt-Schmidt relation for the simulations involved in this study. We use two panels for clarity; in the bottom panel we only plot the MIRAGE sample, while on the top panel we over-plot the MASSIV data on the MIRAGE sample for comparison. In both panels, we also display the relation obtained in \cite{2010ApJ...714L.118D} (red solid line for the relation and dashed line for the associated $1\sigma$ dispersion). Simulations are plotted at different times represented with different colors, with values measured inside the stellar disk scalelength. In case of merger, we ensure to plot only snapshots where the coalescence has been reached. The mergers and isolated disks are respectively plotted with ``+'' and ``$\Delta$''. The MASSIV sample \citep{2012A&amp;A...539A..91C} positions are computed using the half-mass stellar radius for a typical gas fraction of 45\%, and are plotted using black diamonds for isolated galaxies, triangles for minor mergers, and squares for major mergers. The associated error bars are computed using the errors on $H\alpha$ flux, stellar size, and the assumption that the gas fraction $f_g$ lies in the range [0.25,0.65].}
	\label{ks_diagram}
\end{figure*}

{Due to the significant fraction of active galactic nucleus (AGN) in the redshift range $1<z<2$, the inclusion of black-hole particles and the associated AGN feedback should be addressed specifically. Nevertheless, the recent work of \cite{2013MNRAS.434..606G} shows that AGN feedback is unable to disrupt the clumps of 10$^8$-10$^9$ M$_{\odot}$  formed \textit{in-situ} in comparable idealized gas-rich disk simulations. \cite{2013MNRAS.433.3297D} also show that massive clumps may survive AGN feedback during their migration towards the bulge in a fully cosmological context. As the clumps drive most of the SF, we do not expect major changes on short term star formation histories by including AGN feedback. However, at later stages of evolution, the strong heating of the gaseous halo driven by shocks due to AGN feedback should lower the accretion rate, and lead to lower gas fractions in the merger remnants.
Finally, the results obtained in the MIRAGE sample call for further investigations to assess the effect of AGN feedback in such kind of simulations.}

\noindent We have the opportunity with the MIRAGE sample to prospect further questions on galaxy evolution. The combination of statistical probing of the orbital parameters, the controlled input parameters due to idealized framework, the parsec scale resolution, and the explicit physically-motived implementation of stellar feedback make it a consequent database which will be used for studying (i) the properties and lifetime of the giant (10$^8$-10$^9$ M$_{\odot}$) star-forming clumps (already presented in \citealt{2013arXiv1307.7136B}), (ii) the impact of the migration and interaction of the clumps on the galaxies properties, (iii) the metallicity evolution in mergers and isolated disks (iv) {the classification of velocity fields of high-z galaxies based on a large set of mock observations derived from the MIRAGE sample, among other studies.}

\normalfont
\begin{acknowledgements}
We thank the anonymous referee for the useful comments that greatly improved this paper. We thank the whole MASSIV team for encouraging this project with stimulating discussions.
The simulations presented in this work were performed at the \textit{Tres Grand Centre de Calcul} of CEA under GENCI allocations 2012-GEN 2192 and 2013-GEN2192, and at the LRZ SuperMUC facility under PRACE allocation number 50816. FR and FB acknowledge funding form the EC through grant ERC-StG-257720. We thank J. Billing for the distribution of the \verb|starscream| code under the GPL license.
Snapshots format conversions were managed using the UNSIO library (\url{http://projets.lam.fr/projects/unsio}).
\end{acknowledgements}

\bibliography{paper1}

\newpage

\appendix
\section{Simulations maps}
\label{append}
\begin{figure*}[htbp]
\centering
\includegraphics[width=19cm]{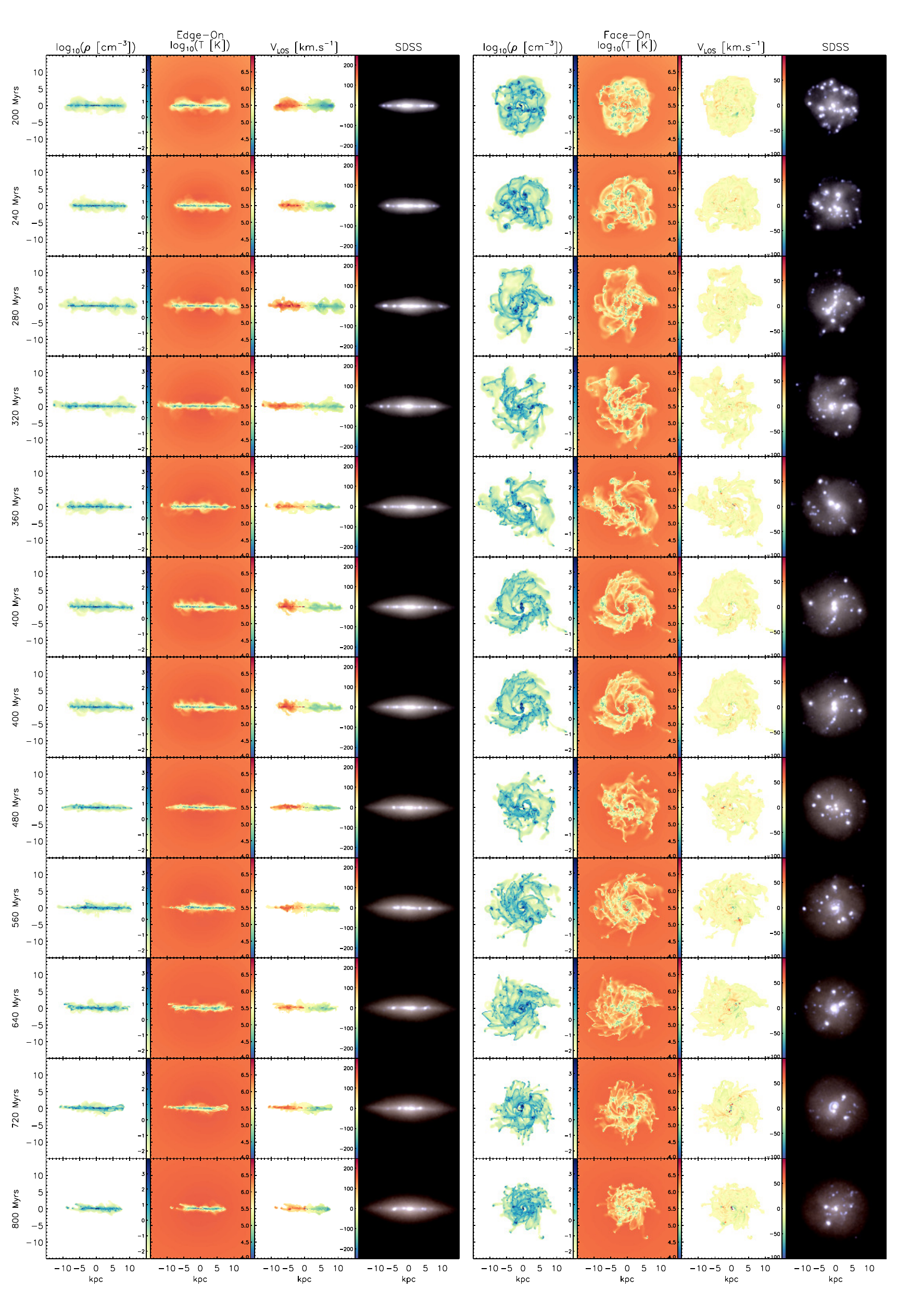}
\caption{Orthographic projections of the G2 simulation. Two distinct perpendicular line-of-sight (LOS) are displayed using two sets of four columns. The edge-on view is displayed on the four left columns, and the face-on view is displayed in the four right columns. The projection angles are kept constant with respect to the original cartesian axes. For each projection, we show the mass-weighted mean gas density (first column), the gas temperature (second column), the mass-weighted mean gas radial velocity (third column), and a mock  SDSS u/g/r  composite image (fourth column). The gas density, gas temperature, and the velocity range are encoded in the inner right side of each figure.}
\label{G2_paper1}
\end{figure*}

\begin{figure*}[htbp]
\centering
\includegraphics[width=19cm]{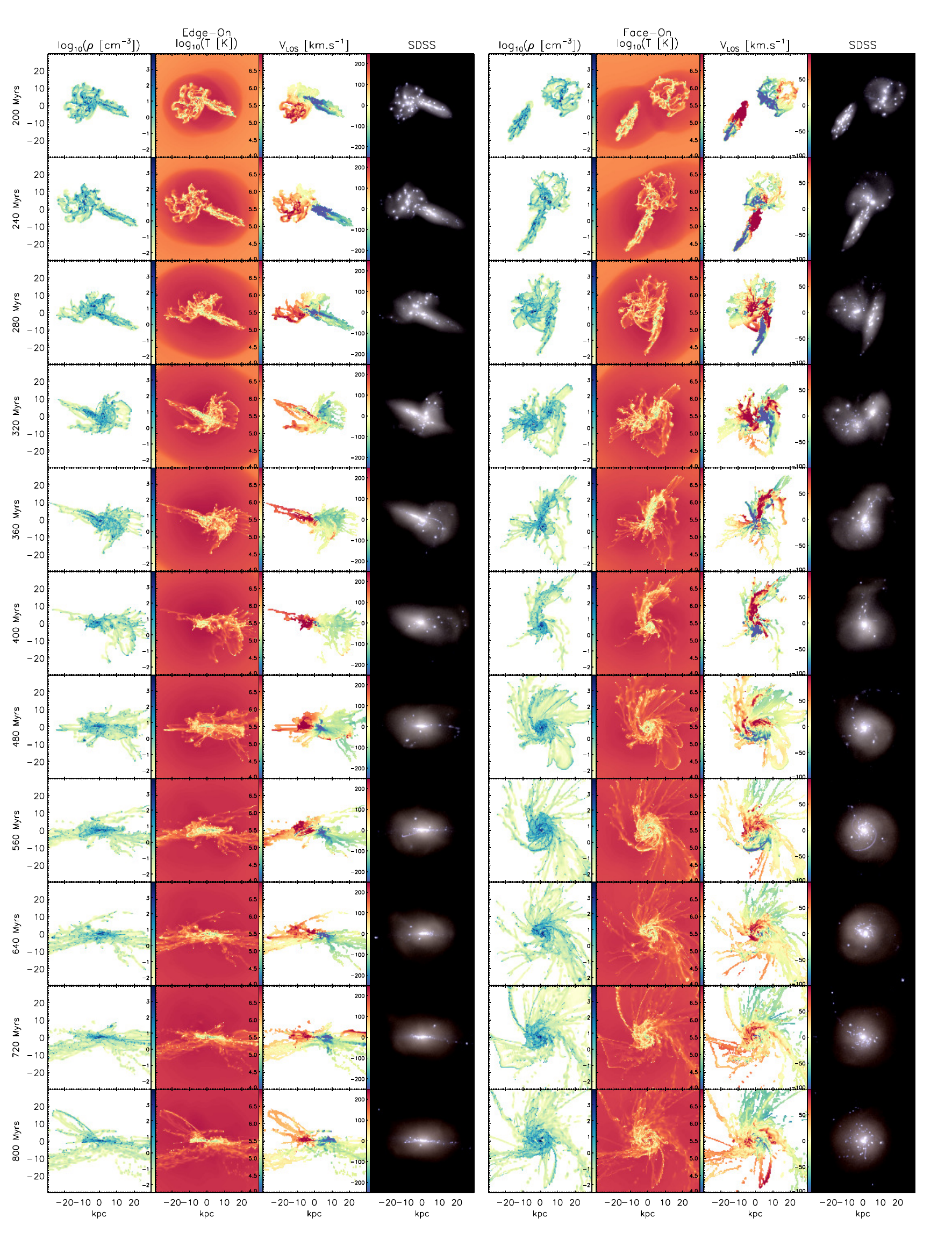}
\caption{Same as figure \ref{G2_paper1} for simulation G1\_G1\_90\_0\_90.}
\label{G1_G1_90_0_90_paper1}
\end{figure*}
 
\begin{figure*}[htbp]
\centering
\includegraphics[width=19cm]{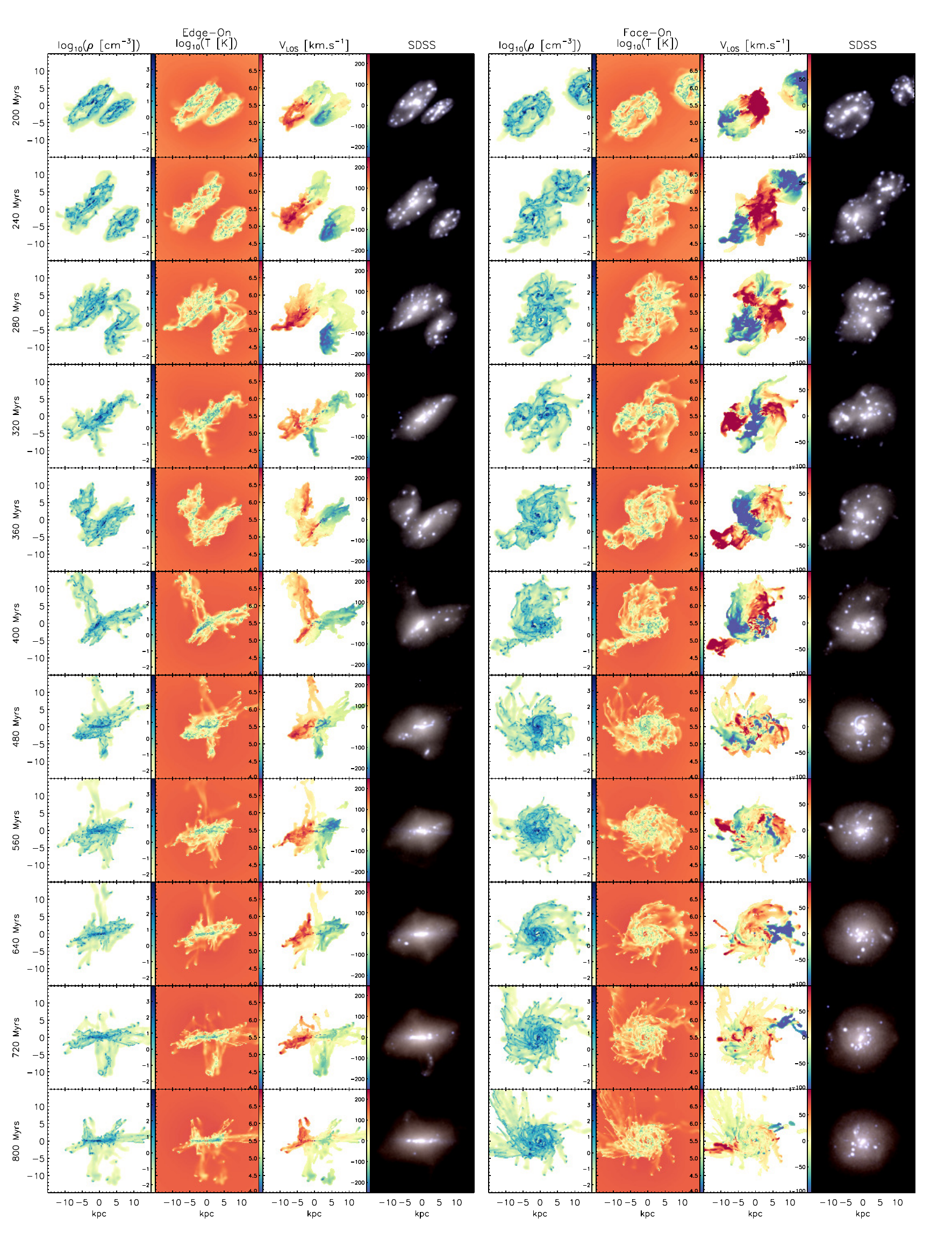}
\caption{Same as figure \ref{G2_paper1} for simulation G2\_G3\_90\_90\_180.}
\label{G2_G3_90_90_180_paper1}
\end{figure*}

\end{document}